\documentclass{pasj01}
\usepackage{lineno}

\begin{document}

\author{Mariko \textsc{Kimura}\altaffilmark{1,*}, and 
        Yoji \textsc{Osaki}\altaffilmark{2}
        }
\email{mariko.kimura@riken.jp}

\altaffiltext{1}{Institute of Physical and Chemical Research (RIKEN), 2-1 Hirosawa,
Wako, Saitama 351-0198}
\altaffiltext{2}{Department of Astronomy, School of Science, University of Tokyo, Hongo, Tokyo 113-0033}

\title{
The light curve simulations of the 2021 anomalous event in SS Cygni
}

\Received{} \Accepted{}


\SetRunningHead{Kimura, M., and Osaki, Y.}{}

\maketitle

\begin{abstract}

The prototype dwarf nova SS Cyg unexpectedly exhibited 
an anomalous event in its light curve in the early few months
of 2021 in which regular dwarf nova-type outbursts stopped,
but small-amplitude fluctuations occurred only.
Inspired by this event, we have performed numerical
simulations of light curves of SS Cyg by varying mass
transfer rates and varying viscosity parameters in
the cool disk.
We have also studied the effect of gas-stream overflows
beyond the outer disk edge in the light curve simulations.
We have confirmed that the enhanced mass transfer is unlikely
responsible for the 2021 anomalous event and its forerunner.
We have found that the enhancement of the viscosity
in the disk may reproduce the forerunner of that event but
may not be enough to explain the 2021 anomalous event,
\textcolor{black}{although the latter result might be particular
to our thermal equilibrium curve used.
Within our simulations, a model of the gas stream overflow with
a slightly higher mass transfer rate than that of our standard 
model reproduces light curves similar to the 2021 anomalous event}.
We suggest that the gas-stream overflow is necessary to 
reproduce that event.
The gas-stream overflow may also be responsible for
the abnormally high X-ray flux during the normal quiescent
state in SS Cyg.

\end{abstract}


\section{Introduction}

Dwarf novae (DNe), one subclass of non-magnetic 
cataclysmic variables (CVs) are eruptive variables 
that show sudden brightening called ``outbursts'' of 
amplitudes with 2--6 mag and with a typical repetition 
cycle of a few weeks to several months.  
CVs are close binary systems composed of the primary white 
dwarf (WD) and the secondary low-mass star.  An accretion 
disk is formed around the WD by the transferred mass from 
the secondary star (see \cite{war95book} for a review).  
The DN outbursts are caused by sudden brightening in 
the accretion disk, which is, in turn, now thought to be 
caused by instabilities in the accretion disk.  
This model is called the disk instability model (DIM),
which was firstly proposed by \citet{osa74DNmodel}.  
In this model, it is considered that the disk stores mass 
provided by the secondary star in the quiescent state, 
and that the accumulated mass accretes onto the WD 
in the outburst state.  
The physical mechanism of this instability was found to 
be the thermal-viscous instability triggered by partial 
ionization of hydrogen \citep{hos79DImodel}.  
The disk experiences relaxation oscillations between 
the hot state with high accretion rates and 
the cool state with low accretion rates (i.e., 
the thermal limit cycle). The numerical simulations have 
been successfully performed along the DIM by the several 
different groups to reproduce the observed light curves 
of DNe (see, e.g., \cite{can93review,osa96review,ham20review} 
for reviews).  

Non-magnetic CVs above the period gap are classified 
into three subclasses: 
SS Cyg-type stars, Z Cam-type stars, and nova-like stars (NLs).  
SS Cyg stars repeat dwarf-nova outbursts, and NLs experience 
no outbursts.  
In the DIM, the critical mass transfer rate ($\dot{M}_{\rm crit}$) 
distinguishes these subclasses.  
If the mass transfer rate is lower than $\dot{M}_{\rm crit}$, 
the disk becomes thermally unstable, and the system repeats 
outbursts, which is recognized as a SS Cyg star.  
If the mass transfer rate is higher than $\dot{M}_{\rm crit}$, 
the disk is in a thermally stable hot state, and the system shows 
constant high luminosity, which is observed as an NL.  
Z Cam stars are the intermediate class between SS Cyg stars 
and NLs, having mass transfer rates close to $\dot{M}_{\rm crit}$.  
They repeat DN outbursts at some time and from time to time 
show a phenomenon called ``standstill'' which is a constant 
luminosity state for some indefinite intervals (months to 
more than 1 yr) as if they are a hybrid class between 
SS Cyg stars and NLs.  
There exists another class of CVs called IW And-type stars 
(or ``anomalous Z Cam-type stars'').  
The standstill in the ordinary Z Cam stars is terminated by 
fading to the quiescent state, while the standstill 
(or, more precisely speaking ``quasi-standstill'') is rather 
terminated by brightening to small-amplitude outbursts. 
IW And stars exhibit a characteristic light variation; 
a repetition of a quasi-standstill (i.e., a mid-brightness 
interval with small-amplitude oscillatory light variations) 
terminated by brightening \citep{kat19iwand}. 

\begin{figure*}[htb]
\begin{center}
\FigureFile(160mm, 50mm){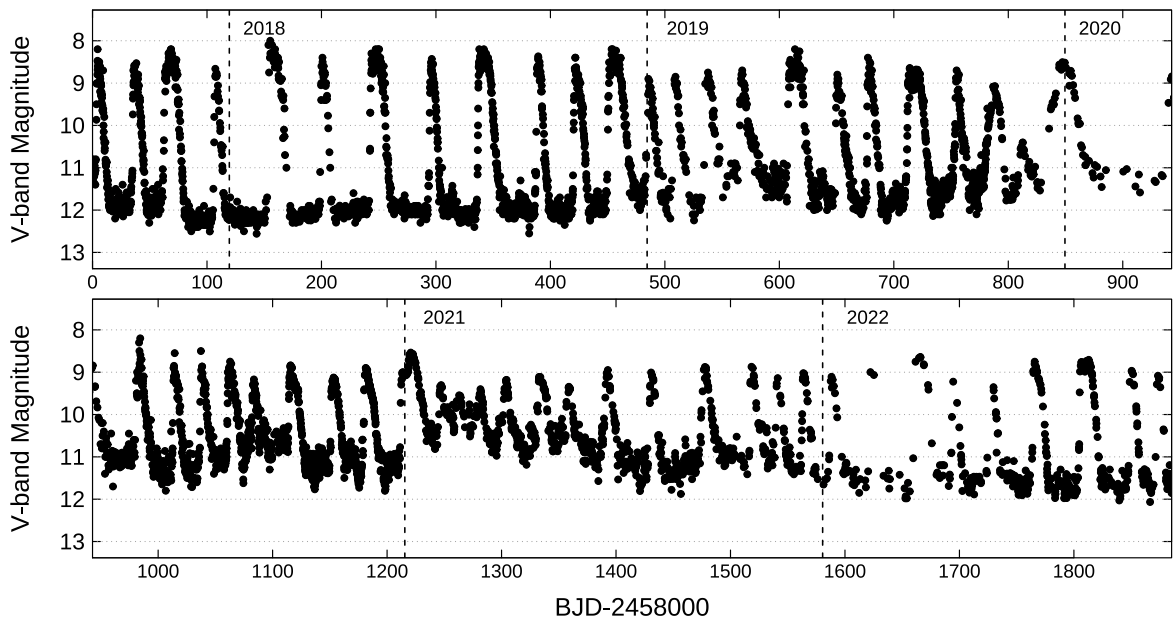}
\end{center}
\caption{
Optical light curve of SS Cyg after BJD 2458000.  The data are taken from the AAVSO archive\footnote{$<$http://www.aavso.org/data/download/$>$}.  We use the $V$-band photometry and the visual observations.  
}
\label{lc-normal}
\end{figure*}

\begin{figure*}[htb]
\begin{center}
\FigureFile(120mm, 50mm){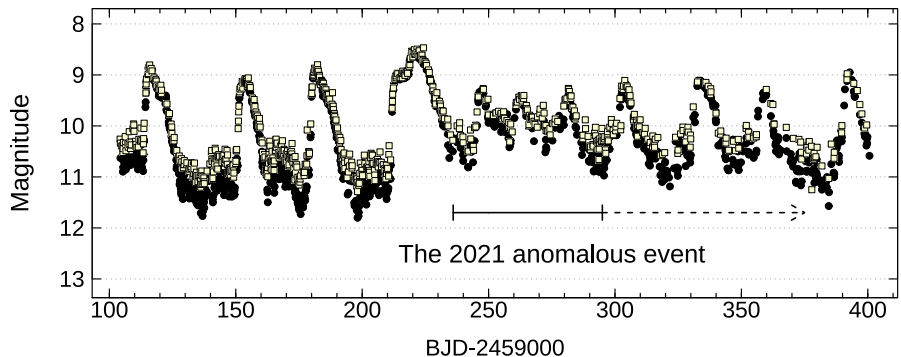}
\end{center}
\caption{
Expanded optical light curve of SS Cyg around the 2021 anomalous event, adopted from Fig.~2 in \citet{kim21sscyg}.  The data are taken from the AAVSO archive\footnote{$<$http://www.aavso.org/data/download/$>$} and the Variable Star Network (VSNET), and the same as that used in \citet{kim21sscyg}.  The filled circles and open squares denote the $V$-band and $R_{\rm C}$-band photometric data, respectively.
}
\label{lc-standstill}
\end{figure*}

The brightest dwarf nova SS Cyg has been monitored 
for more than 100 years by visual observations and 
optical photometry.  
This system has repeated DN outbursts with quiescence 
intervals of $\sim$1 month for a long time 
(see the AAVSO historical light curve\footnote{$<$https://www.aavso.org/historic-light-curves$>$}), and has been recognized 
as the prototype of SS Cyg-type stars.  
However, SS Cyg unexpectedly began to show an anomalous 
light curve, i.e., something like ``standstill'' 
from the end of January 2021 as if it imitated Z Cam stars 
(see, Figures \ref{lc-normal} and \ref{lc-standstill}). 
Figure \ref{lc-normal} exhibits the recent light curve 
of SS Cyg over 5 years beginning from BJD 2458000 
while Figure \ref{lc-standstill} shows the expanded 
one around the anomalous event.  
When this happened in the 2021 early few months, 
we suspected it could be something like 
a standstill in Z Cam stars because the brightness 
of the star stayed more or less constant and the mean 
light level increased to $\sim$10 mag (vsnet-alert 25453).  
However, it soon turned out that the anomalous event 
was not a genuine standstill, but the system showed rapid 
oscillatory light variations with small amplitude 
in which the maxima and minima seemed to approach each other. 
We call this event ``the 2021 anomalous event 
in SS Cyg'' in this paper, a phrase already used 
in \citet{kim21sscyg} (Event B there).  
The X-ray luminosity during the anomalous event was 
$\sim$3$\times$10$^{33}$~erg~s$^{-1}$, which is 
about 5-times higher than that in the normal quiescence, 
and something unusual happened over wide wavelengths 
(see, \cite{kim21sscyg}).  
Such an event had never been observed in SS Cyg in 
its long history of observations.  

Let us scrutinize the overall light curve of SS Cyg 
extending nearly five years shown in Figure \ref{lc-normal}.   
In its earliest phase, the light curve of the year 2018, 
in particular, was that of the typical one in SS Cyg 
in which eight outbursts were observed in one year with 
the minimum around 12 mag and the maximum reaching 8.5 mag, 
so that the outburst amplitude was $\sim$3.5 mag and 
the long and short outbursts alternated.  
However, something unusual began to occur already in 
2019--2020. 
The optical and X-ray quiescent levels simultaneously 
and gradually rose since around August 2019 (see also 
\cite{kim21sscyg}).  
One of the authors (M.~Kimura) noticed that and requested 
intensive optical photometry of SS Cyg to the AAVSO team 
\citep{waa20sscyg} in September 2020.  
Although nobody knew what was going on in SS Cyg at that time, 
this phenomenon in 2019--2020 could be a forerunner of 
the anomalous event in 2021. 
As mentioned above, the 2021 anomalous event then started 
to occur from the end of January 2021.  

However, we can see from Figure \ref{lc-normal} that 
the 2021 anomalous event was short-lived (for a couple 
of months) and the system seemed to return gradually to 
the normal state.  
After about 60 days from the beginning of the anomalous event, 
the outburst amplitude became gradually larger and 
the outburst frequency became gradually lower.  
The X-ray luminosity was also decreasing.  
The main part of the anomalous event seems to be the first 
60~d after the long outburst in 2021 (see the horizontal 
solid-line bar in Figure \ref{lc-standstill}) with 
saw-tooth-like light variations (rapid-rise, slow-decline type) 
of amplitude 1~mag, and 
the anomaly of this event was gradually fading after 
this part, though we cannot determine exactly when this event ended.  
Currently (in November 2022), SS Cyg seems to have 
returned back to the normal state (see the upper panel of 
Figure \ref{lc-normal}).
We can see that the quiescent flux showed wavy modulations 
on long timescales of $\sim$100~d (see Figure 
\ref{lc-normal}).  A general tendency is that  
the brighter the quiescent level is, 
the lower the outburst amplitude is, the higher 
the outburst frequency is.  
The 2021 anomalous event occurred when the quiescent level 
was maximized in these undulations.

Motivated by the 2021 anomalous event,
we started performing numerical simulations of 
the light curve of SS Cyg to reproduce this anomalous event.  
Since we initially thought that the 2021 anomalous event 
could be a genuine standstill of Z Cam stars, 
we tried, for instance, to increase the mass transfer rate 
up to the critical mass transfer rate. 
However, this anomalous event was not a genuine standstill. 
Nevertheless, this event in itself is definitely unusual 
and it will undoubtedly provide 
a great opportunity to reconsider the nature of 
standstill and/or standstill-like phenomena in DNe.  
In this paper, we have extensively performed the light curve 
simulations of SS Cyg by taking this anomalous event in mind.
To do so, we examine the response of the light curve with 
varying mass transfer rates and the varying viscosity parameter 
of $\alpha$.  
\textcolor{black}{
In the previous observational paper by \citet{kim21sscyg}, 
they suggested that enhancement of viscosity in the disk 
could cause the anomalous event and its forerunner.
They suggested possible reasons for the enhancement of 
viscosity in section 4.2 of that paper.
Another possibility may be the effect of gas stream overflow 
because the gas stream overflow can keep the inner disk 
in the hot state as discussed by \citet{kim20kic940}.
We, therefore, study the effect of the possible gas stream 
overflow beyond the outer disk rim.
Although the reason for the sudden occurrence of the stream 
overflow is unknown, the change in the thermal state of 
the outer disk and vertical oscillations of the disk induced 
by the collision of the stream with the disk might cause 
the overflow \citep{hes99streamoutflow}.
}

\citet{can93DI} has already studied the light curve 
response of SS Cyg to varying input parameters, in particular, 
on the mass transfer rate and the viscosity parameter, 
extensively.
Since such an anomalous event was not expected at that time, 
the parameter ranges in his simulations were not wide enough 
to examine the present anomalous event.  
Furthermore, the treatments of the outer disk radius, 
tidal torque, and tidal dissipation are different 
between ours and those of \citet{can93DI}; we treat variable 
outer disk radius in our simulations, while he used the fixed 
disk radius.  
The variable disk radius and the treatment of tidal torques 
and of the tidal dissipation play the essential role 
in our case.  

This paper is structured as follows.  
In section~2, we describe the method of our numerical 
simulations and set the standard model parameters for SS Cyg.  
In section~3, we present the simulation results for varying 
mass transfer rates, varying viscosity parameters, and 
varying overflow rates in relation to the 2021 anomalous 
event in SS Cyg.  
Finally, we discuss our results in section~4 and summarize 
our findings in section~5.

\section{Method of numerical simulations and standard model parameters}

In this paper, we perform numerical simulations for 
the time evolution of an accretion disk to study 
outburst light curves in CVs.  
The method of our numerical simulations is basically 
the same as that described in 
\citet{ich92diskradius} and \citet{kim20tiltdiskmodel}. 
We calculate the time evolution of a geometrically thin and 
axisymmetric disk (but ``non-tilted'' in this paper).  
We adopt the one-zone approximation in the vertical direction,
by assuming a geometrically-thin disk in hydrostatic 
equilibrium. 
We formulate mass, angular momentum, and energy conservation 
laws for the accretion disk in the radial direction, 
and solve them by a hybrid method of explicit and implicit 
integration.  
We make some minor modifications to the numerical code 
of \citet{kim20tiltdiskmodel},   
the details of which are described in Appendix 1. 
The disk radius is time-varying in our formulation to conserve 
the total angular momentum of the disk.  

In this study, we assume that the tidal truncation radius 
works as a solid brick wall.  
When the disk reaches the tidal truncation radius and tries 
to expand further, the expansion is stopped at the tidal 
truncation radius by the strong tidal torques and the disk 
cannot expand further.  
In our previous paper (\cite{kim20tiltdiskmodel}), 
the removal of the extra angular momentum was taken into 
account, but the extra tidal heating was not considered.  
In this paper, we take into account the extra tidal heating 
at the outer disk edge due to the tidal truncation 
as one of the heating functions.  
Its formulation and effect on the simulations are 
presented in Appendix 2.

The binary parameters used in this paper for SS Cyg are as follows: 
the orbital period ($P_{\rm orb}$) is 0.27512973~d, 
the WD mass ($M_1$) is 0.94$M_{\solar}$, 
the mass of the secondary star ($M_2$) is 0.59$M_{\solar}$, 
the binary separation ($a$) is $1.427 \times 10^{11}$~cm,
the inclination angle ($i$) is 45~deg, 
the tidal truncation radius ($r_{\rm tidal}$) is 
0.356$a$ = $5.08 \times 10^{10}$~cm, 
the Lubow-Shu radius (or the circularization radius, $r_{\rm LS}$) 
is 0.105$a$ = $1.50 \times 10^{10}$~cm,  
respectively \citep{pac77diskmodel,hes84sscyg,hil17sscyg}.  
In calculating the $V$-band magnitude, we use the distance 
of SS Cyg, which is 114.25~pc measured by the {\it Gaia} parallax 
\citep{bai18gaia}.  
We also add the brightness of the Roche-lobe filling secondary 
star with the temperature of 4,200~K, which corresponds to 
12.8~mag, and the brightness of the WD of 50,000~K, which 
corresponds to 15.3~mag (see, \citet{ham20modeling}).  
We regard the flux from these two objects as constant 
for simplicity.  

One of the most important factors for simulations 
of dwarf-nova outbursts is the selection of viscosity 
parameter of \citet{sha73BHbinary}, $\alpha$'s: 
$\alpha_{\rm hot}$ for the hot disk in outburst, and 
$\alpha_{\rm cool}$ for the cool disk in quiescence,
which affects the `S'-shaped thermal equilibrium curve 
characterizing the outburst behavior. 
If the ratio $\alpha_{\rm hot} / \alpha_{\rm cool}$ is 
larger, the outburst amplitude becomes larger 
\citep{sma84DI,min85DNDI}.  
The viscous timescale is longer for smaller $\alpha$, 
so that the outburst duration becomes longer as 
$\alpha_{\rm hot}$ decreases, and the recurrence time of 
outbursts increases as $\alpha_{\rm cool}$ decreases.  
We have performed the parameter search to reproduce 
the characteristics of the typical observed light curve 
in SS Cyg by our code.
The parameters adopted here for our standard model for 
SS Cyg are $\dot{M}_{\rm tr} = 10^{16.9}$~g~s$^{-1}$, 
$\alpha_{\rm hot}$ = 0.15, and 
$\alpha_{\rm cool}/\alpha_{\rm hot} = 0.1$, 
in order to reproduce the outburst interval (the quiescence 
duration) of $\sim$20--40 days, the outburst amplitude of 
$\sim$4 mag, and the duration of outbursts of $\sim$10--20~days, 
respectively.  The radial dependence of $\alpha_{\rm cool}$ is 
not implemented in this paper because it is not necessary 
in the case of reproducing normal outbursts in SS Cyg, 
while we did in \citet{kim20tiltdiskmodel}.  

\begin{figure*}[htb]
\begin{center}
\begin{minipage}{0.49\hsize}
\FigureFile(80mm, 50mm){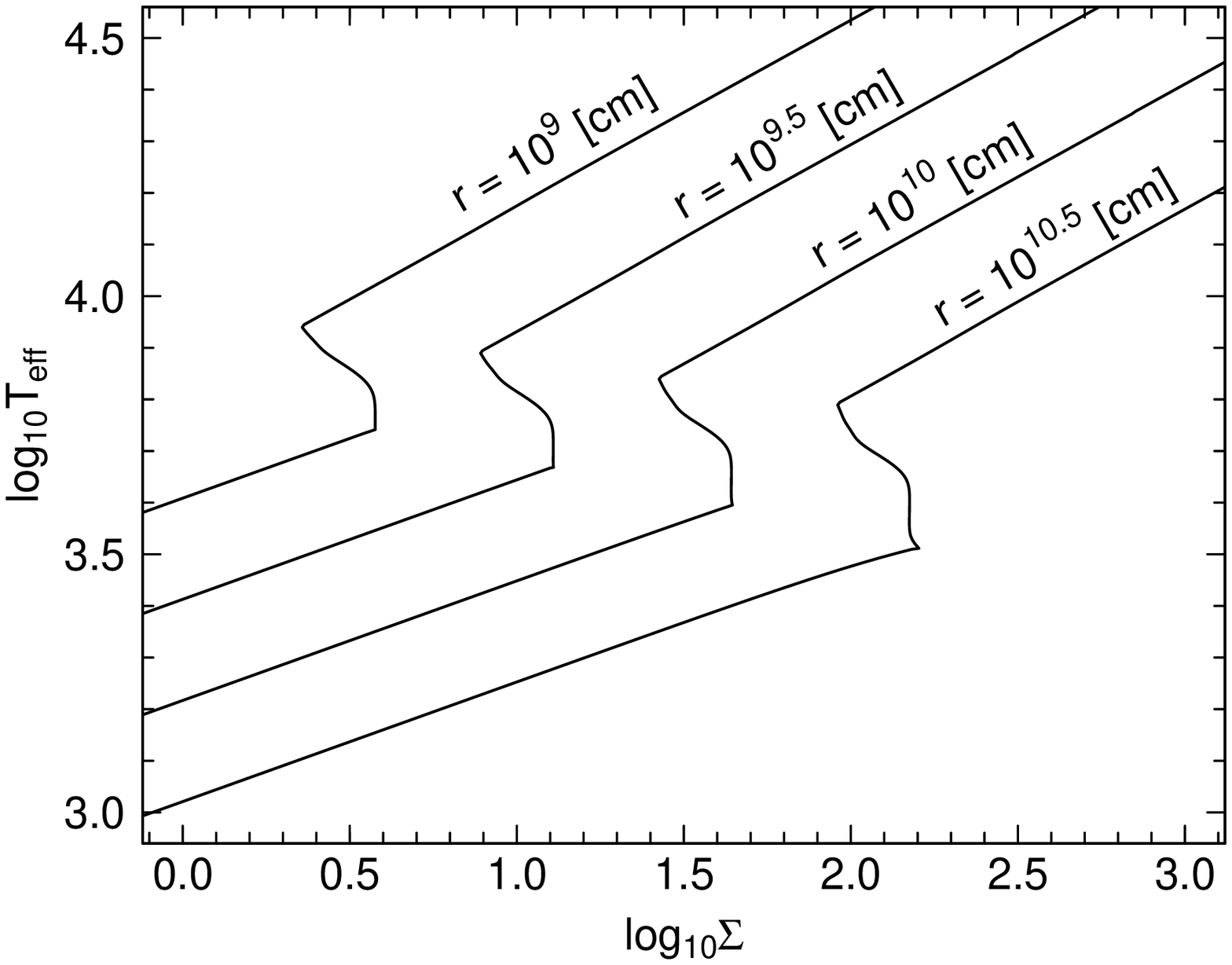}
\end{minipage}
\begin{minipage}{0.49\hsize}
\FigureFile(80mm, 50mm){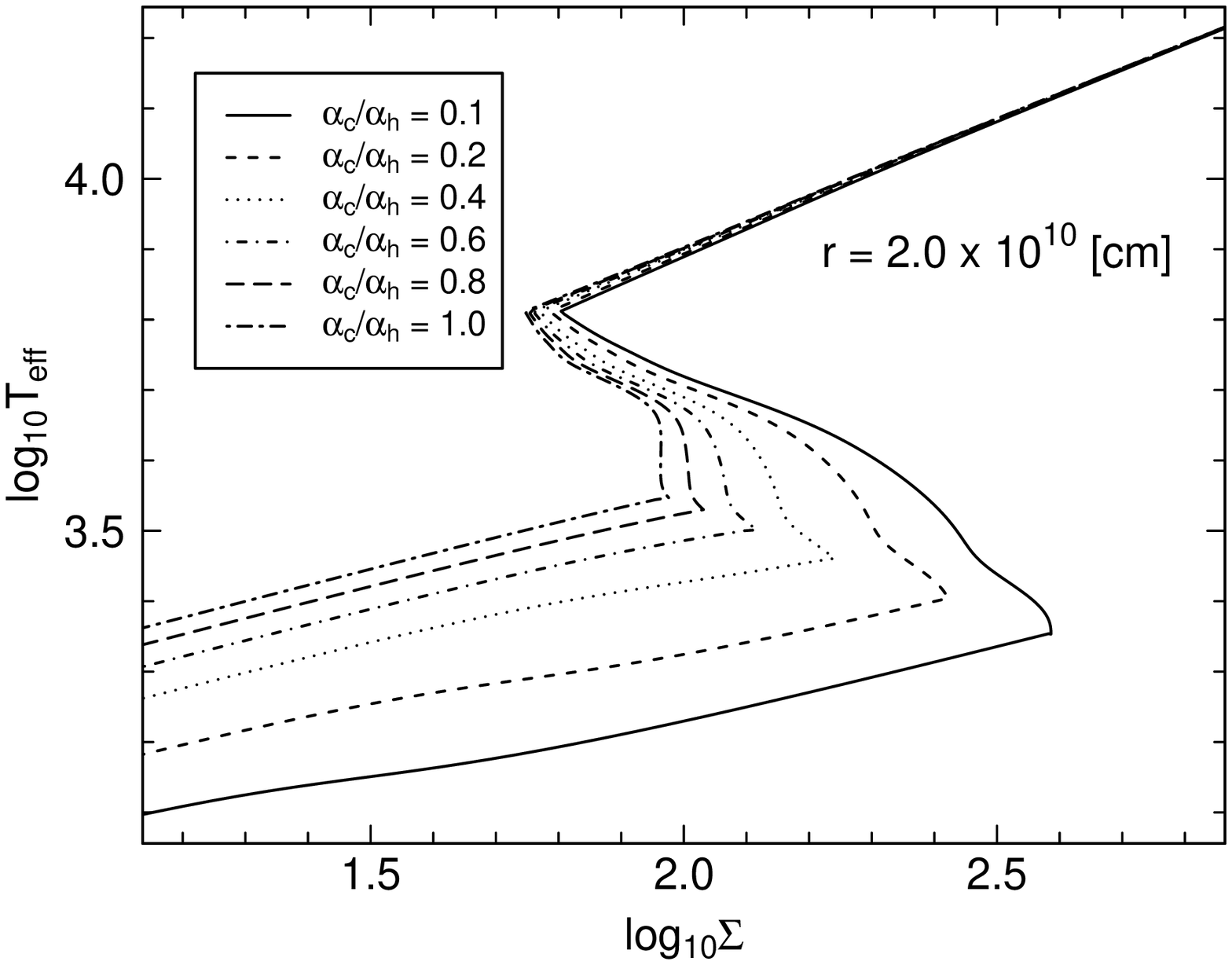}
\end{minipage}
\end{center}
\caption{
(Left) Thermal equilibrium curves at $r$ = $10^{9}$, $10^{9.5}$, $10^{10}$, and $10^{10.5}$~cm for our basic parameter sets in section 2.  Here, $\alpha_{\rm hot}$ = 0.15 and $\alpha_{\rm cool} / \alpha_{\rm hot}$ = 1.0.  
(Right) Comparison of thermal equilibrium curves at $r$ = $2.0 \times 10^{10}$~cm for $\alpha_{\rm hot}$ = 0.15 with different $\alpha_{\rm cool} / \alpha_{\rm hot}$.  
These equilibrium curves are calculated by $2F = Q_1^{+}$ (see \citet{kim20tiltdiskmodel} for the expressions of $F$ and $Q_1^{+}$).  
}
\label{scurves}
\end{figure*}

\textcolor{black}{
The `S'-shaped thermal equilibrium curve that we use 
in this paper is calculated on the basis of Fig.~3 
of \citet{kim20tiltdiskmodel}, which is a simple 
approximate expression for Fig.~2 of \citet{min83DNDI}.}  
The left panel of Figure \ref{scurves} presents examples 
of thermal equilibrium curves for four representative annuli 
of the disk with $\alpha_{\rm hot}$ = 0.15 and 
$\alpha_{\rm cool}/\alpha_{\rm hot} = 1.0$.  
We exhibit in the right panel of the same figure the thermal 
equilibrium curves at $r$ = $2.0 \times 10^{10}$~cm with 
various $\alpha_{\rm cool} / \alpha_{\rm hot}$ values.  
As $\alpha_{\rm cool} / \alpha_{\rm hot}$ is increased, 
the cool branch rises upwards in the vertical direction, 
the critical surface density $\Sigma_{\rm max}$ above 
which there is no stable solution in the cool branch 
decreases, and the unstable intermediate branch becomes 
shorter.

\textcolor{black}{
We note here that different authors used different thermal 
equilibrium curves in past studies 
since different authors calculated the vertical structure 
of the disk in different ways, such as in the treatment of 
convective energy transport and opacities used, and so on. 
Our thermal equilibrium curve is based on the computations 
in \citet{min83DNDI} in which the vertical structure 
was solved, and the convective energy transport was taken into 
account by the mixing length theory, but molecular opacities 
were not considered. 
As seen in Fig.~3 and Fig.~14 of \citet{min83DNDI}, 
the thermal equilibrium curve takes $\xi$-shaped curve and 
has two unstable branches (i.e., two local maxima in $\Sigma$ 
in the cold branch).
As discussed by \citet{can93DIreview}, 
the lower maximum is determined by the opacity, while 
the upper maximum is determined by convection (see Fig.~2 
of \citet{poj86structure}).
The end of the cool branch in our thermal equilibrium curve 
is the lower maximum of the two. 
As seen in Fig.~14 of \citet{min83DNDI}, the lower maximum 
in $\Sigma$ is dominant when the convective energy transport 
is inefficient, while the upper maximum is dominant 
when the convective energy transport is very efficient. 
For example, the thermal equilibrium curve used in 
\citet{ham98diskmodel} corresponds to the case of very 
effective convection (see Figure 1 of that paper). 
In their thermal equilibrium curve, the maximum effective 
temperature in the cold branch is independent of 
$\alpha_{\rm cool}/\alpha_{\rm hot}$, which is higher 
than ours.
The main source of the opacity in the cool branch 
of our thermal equilibrium curve is negative hydrogen, 
and $\Sigma_{\rm max}$ in the cool branch of our thermal 
equilibrium curve corresponds to the point at which 
the disk changes from the optically thin state to 
the optically thick state, and there is still controversy 
whether the disk in the cold state is optically thin or thick.
Molecular opacities were not considered in our thermal 
equilibrium curve, a certain drawback in our calculations. 
However, molecular opacity is important if the temperature 
is lower than $\sim$2,500~K (see Fig.~2 of 
\citet{can84ADvertical}). 
Our thermal equilibrium curve thus tends to have the maximum 
effective temperature of the cool disk lower than 
those in other works such as \citet{ham98diskmodel}. 
}

The inner edge of the disk, $r_{\rm in}$, is also 
one of the important parameters.  
In the quiescent state, it is thought that 
the inner disk is truncated via evaporation of 
the disk mass by the coronal siphon flow from 
theoretical and observational studies 
\citep{mey94siphonflow,bal12xrayDNe}.  
In this case, if $r_{\rm in}$ becomes larger, 
inside-out outbursts are suppressed, and the X-ray flux 
in quiescence increases \citep{can93DI}.  
On the other hand, when the outburst occurs, the inner hole 
is thought to be refilled by the accretion flow and 
the inner edge of the disk is thought to extend down to 
the WD surface during outburst.  
In our standard model, we adopt $r_{\rm in}$ to be 
$5 \times 10^{8}$~cm, 
which is approximately the WD radius for the WD mass 
of 0.94$M_{\solar}$ 
\citep{nau72WDmassradius,pro98WDmassradius}.  

We divide the accretion disk between $r_{\rm in}$ and 
the outer disk edge into $N$ concentric annuli 
for our finite difference scheme, where $N$ is variable 
in time because the disk radius is variable in 
our formulations.  
The radial distribution of meshes is equally spaced in 
$\sqrt{r}$ and the maximum number of concentric annuli, 
denoted here by $N_0$, is 200, which occurs when 
the disk outer edge reaches the tidal truncation radius. 

\begin{figure}[htb]
\begin{center}
\FigureFile(80mm, 50mm){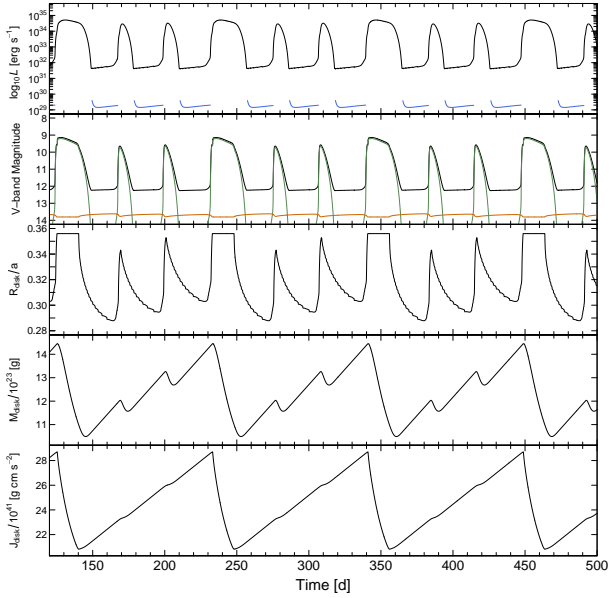}
\end{center}
\caption{
Results of our simulations for $\dot{M}_{\rm tr} = 10^{16.9}$~g~s$^{-1}$, $\alpha_{\rm hot} = 0.15$, $\alpha_{\rm cool}/\alpha_{\rm hot} = 0.1$, and $r_{\rm in} = 5 \times 10^{8}$~cm.  
From top to bottom: the light curve for the bolometric luminosity of the system, that for the $V$-band magnitude, the disk radius in units of the binary separation, the total disk mass, and the total disk angular momentum. 
The blue line in the top panel approximately represents the expected X-ray flux in 
quiescence where $r_{\rm in} = r_{\rm WD}$.
The green and orange lines in the second panel represent the $V$-band magnitude of the disk and the bright spot, respectively.  
The black line in the same panel stands for a sum of the flux from these two components plus the WD and the secondary star.  
}
\label{basic}
\end{figure}

The results of our numerical simulations for normal 
outbursts in SS Cyg by using the parameter sets 
as described above are displayed in Figure \ref{basic}.  
The $V$-band magnitude is estimated by the method 
in \citet{dub18DItest} under the assumption 
that the radiation from the disk is multi-temperature 
blackbody emission.  
The flux from the bright spot is included in the same manner 
as described in \citet{kim20tiltdiskmodel}.  
We see that the recurrence time, the outburst interval, 
the outburst duration, and the repetition of a short 
outburst and a long outburst are consistent with 
the observations of SS Cyg \citep{can92sscyg}.  
The cycle length of long outbursts is $\sim$100~d in 
observations, and this is sensitive to the number of 
short outbursts between two long outbursts.  
In the case of $\dot{M}_{\rm tr} = 10^{16.9}$~g~s$^{-1}$, 
the average cycle length of long outbursts in 
Figure \ref{basic} is $\sim$100~d, which is consistent 
with observations.  
In quiescence, radiation from the secondary star 
and that from the hot spot dominates over that of 
the disk, in our case, just in the same way as the other 
simulations by \citet{can93DI} and \citet{ham20modeling}.  
We note here that the shoulder-like feature (or the precursor) 
in the long outburst is barely visible in the $V$-band light curve.  
This feature is produced when the disk's outer radius reaches 
the tidal truncation radius, and the tidal torques and tidal 
dissipation are greatly increased.  
We regard this model as the standard model in our simulations. 
We perform numerical simulations by changing various parameters 
from the standard model in the following sections.  

\begin{figure*}[htb]
\begin{center}
\begin{minipage}{0.49\hsize}
\FigureFile(80mm, 50mm){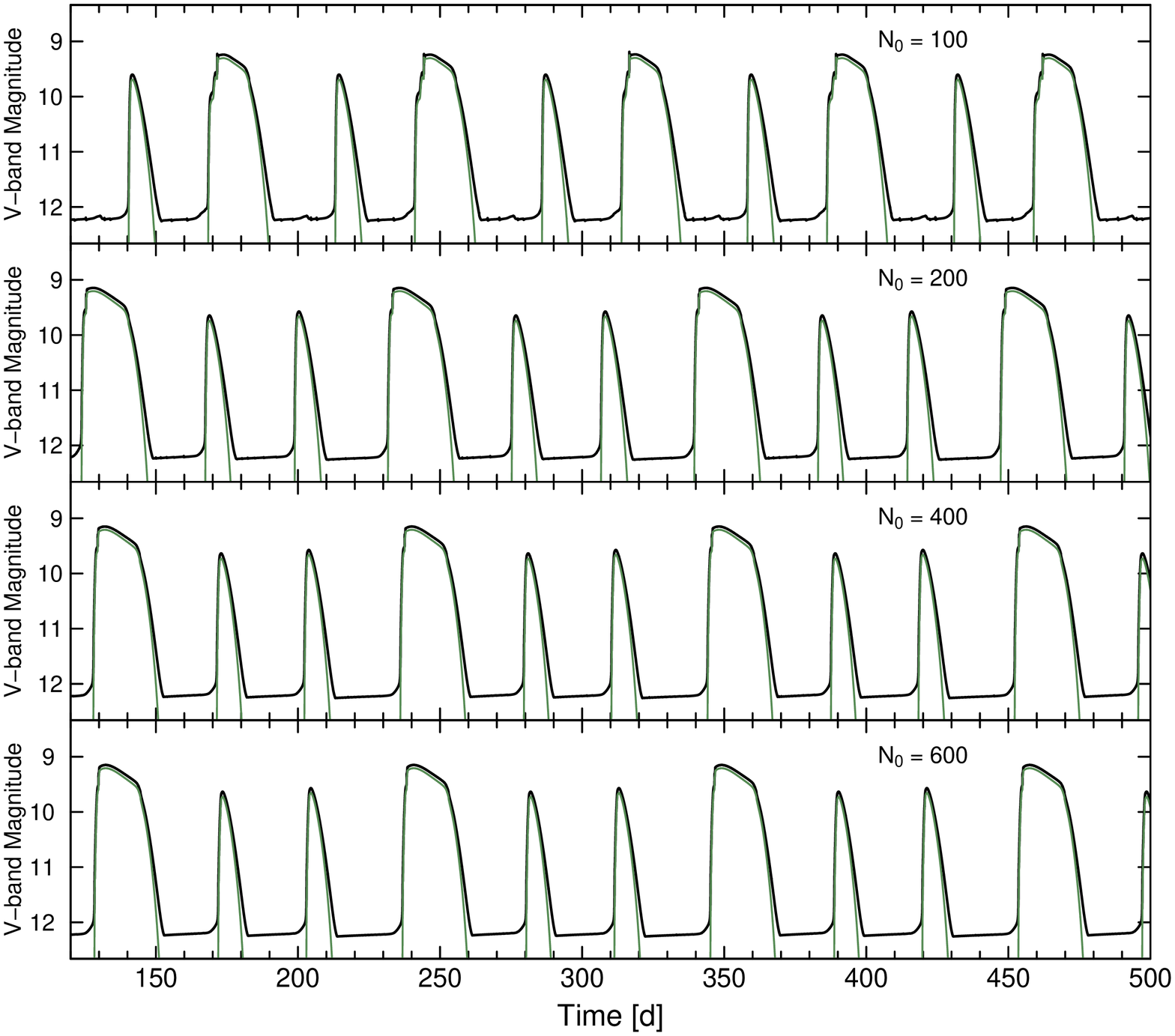}
\end{minipage}
\begin{minipage}{0.49\hsize}
\FigureFile(80mm, 50mm){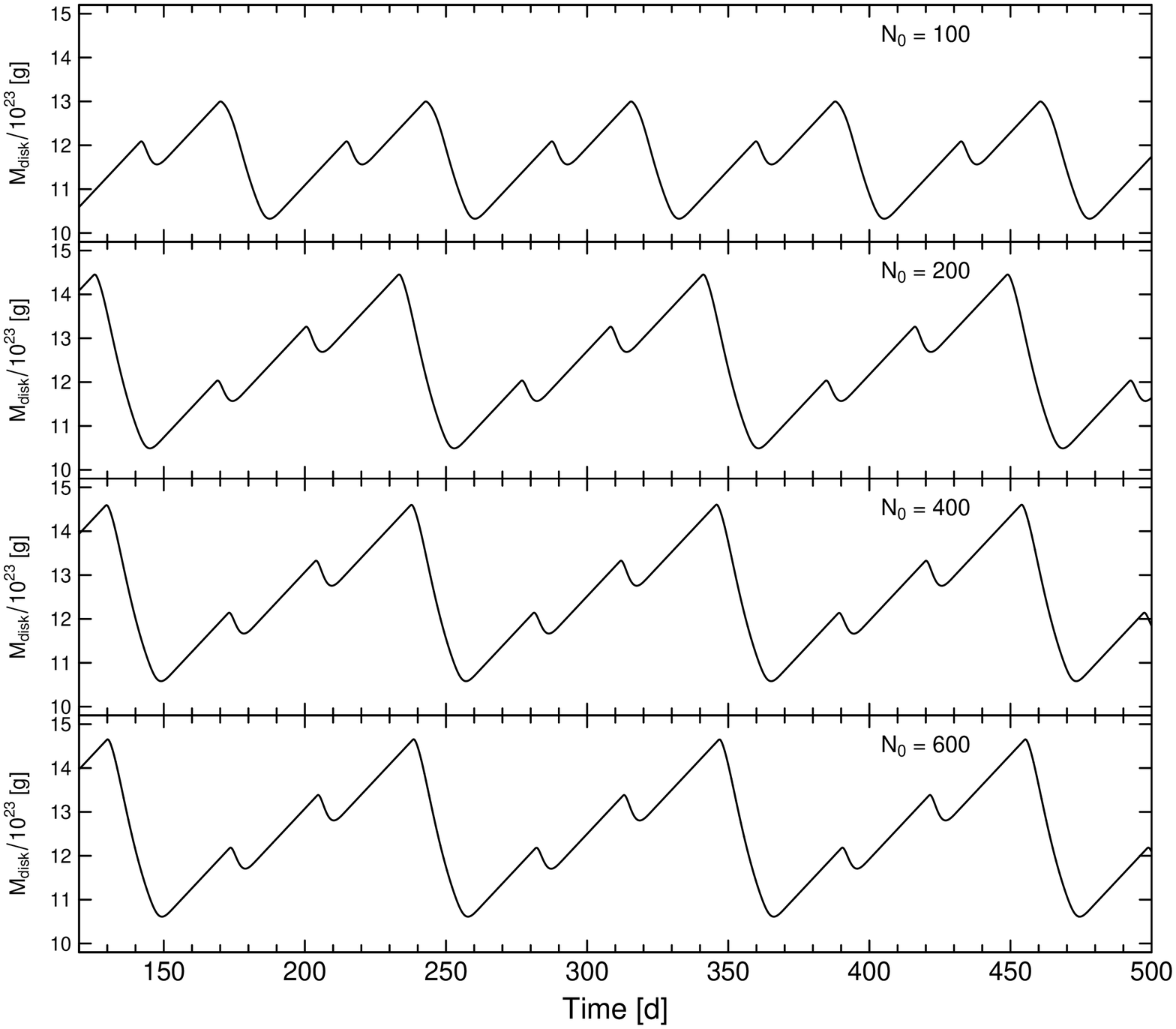}
\end{minipage}
\end{center}
\caption{
(Left) $V$-band light curves of the system and (Right) time evolution of the disk mass at various resolutions.  
The green line in the left panel represents the disk brightness.  
}
\label{grids-check}
\end{figure*}

We have checked if the numerical resolution is enough 
in the case of $N_0 = 200$ by performing simulations 
with coarser or finer grid numbers.  
We display in Figure \ref{grids-check} the $V$-band 
light curves and the time evolution of the disk mass of 
our simulations with $N_0$ = 100, 200, 400, and 600.  
We see that the number of short outbursts between 
two long outbursts is 1 in the case of a coarse grid 
with $N_0 = 100$ (the top panels of Figure \ref{grids-check}), 
but it becomes 2 in the three simulations for finer grids 
with $N_0 \geq 200$.  
The light curve and the time evolution of the disk mass 
are almost the same and indistinguishable among these 
three simulations (see the lower three panels of 
each figure in Figure \ref{grids-check}).  
This result signifies that a resolution with $N_0 \geq$ 200 
is required but the grid number $N_0 = 200$ is enough to 
investigate simulation results.

The X-ray luminosity in quiescence is estimated based on the 
evaporation model.  In this model, the geometrically-thin disk 
is truncated at some inner radius denoted by $r_{\rm evap}$ 
below which the disk expands to the hot spherical corona and 
the central WD is surrounded by the optically thin 
hot corona with a temperature of $\sim 10^7$~K.  
The X-ray luminosity is then given by 
\begin{equation}
L_{\rm X} = G M_1 \dot{M}_{\rm acc} \left( \frac{1}{r_{\rm WD}} - \frac{1}{2r_{\rm evap}} \right),
\label{xray-lumi}
\end{equation}
where $\dot{M}_{\rm acc}$ is the mass supply rate 
and $r_{\rm WD}$ is the WD radius.  
The mass supply rate to the hot corona is estimated by 
the mass accretion rate at $r_{\rm evap}$.  
In this paper, we assume that $r_{\rm evap}$ is equal to 
$r_{\rm in}$.  
Here, we assume that the WD spin is much slower than 
the Keplerian velocity and that the accreted gas does not 
convey any energy inside the WD by dissipating all energy 
in the hot corona.  
Since the local accretion rate at a given radius, $r$, 
is expressed as $\dot{M}_{\rm acc}(r) = 2 \pi r \Sigma (-v_r)$, 
where $\Sigma$ is the surface density, $v_r$ is the radial 
velocity by the viscous diffusion, and $(-v_r) \sim 
\alpha_{\rm cool} (h/r) c_s$, the local accretion rate 
in quiescence at a given radius, $\dot{M}_{\rm acc}(r)$, 
is estimated roughly proportional to $\alpha_{\rm cool} 
T_{\rm c} r^{2.5}$. 
The X-ray luminosity is highly dependent on the inner 
truncation radius, $r_{\rm evap}$ = $r_{\rm in}$, and 
$\alpha_{\rm cool}$.  

We can estimate X-ray luminosity by using simulated mass 
accretion rates at the inner disk edge 
($\dot{M}_{\rm acc}(r_{\rm in})$).  
The blue line in the top panel of Figure \ref{basic} 
approximately represents the X-ray luminosity in 
quiescence where $r_{\rm in} = r_{\rm WD}$.  
The estimated value is $\sim$4$\times$10$^{29}$~erg~s$^{-1}$ 
on average, which is about three orders of magnitude lower 
than the observed X-ray luminosity $\sim$6$\times 
10^{32}$~erg~s$^{-1}$ \citep{whe03sscyg,ish09sscygSuzaku}.  
This discrepancy between the observed and theoretically-predicted 
X-ray luminosity was already pointed out by \citet{whe03sscyg}.  
To increase the X-ray luminosity in the optical quiescence 
in model calculations, one might increase either 
$\alpha_{\rm cool}$ or $r_{\rm in}$, or both.  
Although $r_{\rm in}$ is constant in our simulations, 
the inner disk would be truncated during quiescence in SS Cyg.  
For instance, \citet{bal12xrayDNe} suggested that 
$r_{\rm in}$ is $\sim$5$\times 10^{9}$~cm by their 
observations.  
We have tried simulations with $r_{\rm in} = 5 \times 
10^{9}$~cm and estimated the X-ray luminosity to be 
$\sim$5$\times$10$^{31}$~erg~s$^{-1}$ during quiescence.  
This is, however, still less than a tenth of the observational 
value.

\section{Simulations for the 2021 anomalous event in SS Cyg}

\subsection{Enhanced mass transfer model}

\begin{figure*}[htb]
\begin{center}
\FigureFile(160mm, 50mm){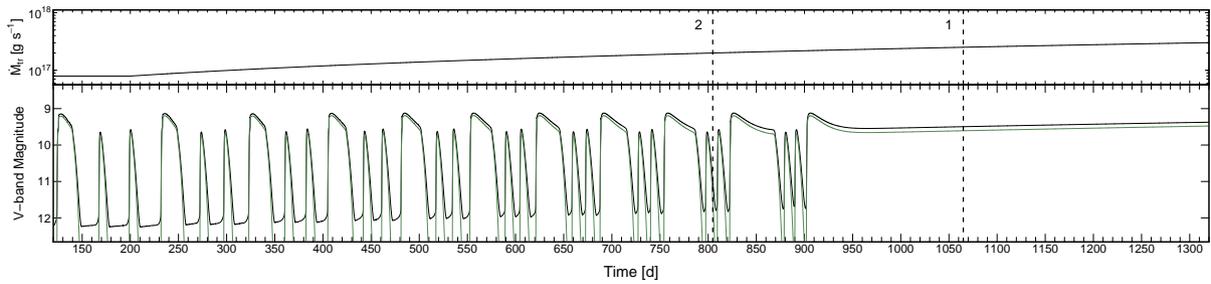}
\end{center}
\caption{
Results of our simulations in which the mass transfer rate is gradually increased with time.
The upper and lower panels represent the mass-transfer rate and the $V$-band magnitude light curve, respectively.
See the caption of Figure \ref{basic} as for the black and green lines in the lower panel.  
}
\label{MTBmodel-increase}
\end{figure*}

We here explore the possibility of enhanced mass transfer 
as a cause for the 2021 anomalous event in SS Cyg.  
As mentioned in the introduction, we originally thought 
that this anomalous event could be a kind of the standstill 
phenomenon in Z Cam stars. 
It is widely believed that the Z Cam-type standstill is produced 
by fluctuations in mass transfer rates where the mass transfer 
rate is enhanced above the critical one above which 
the accretion disk is in the hot stable state 
\citep{mey83zcam,lin85CValphadisk,bua01zcam}.  

We firstly perform simulations by gradually increasing 
the mass transfer rate from that of the standard model 
in order to see what happens in such a case. 
Figure \ref{MTBmodel-increase} illustrates our result of 
simulations.
In Figure \ref{MTBmodel-increase}, the mass transfer rate is  
increased after 200~d with a function of $\dot{M}_{\rm tr} = 
10^{16.9} (1 + 0.0025 (t - 200))$, where $t$ is the time 
in units of days.
We see from Figure \ref{MTBmodel-increase} 
that the cycle length of outbursts decreases with 
the increase in mass transfer rates until $\sim$800~d, 
but it begins to increase after that, mainly 
because the duration of the long outburst becomes longer.  
Eventually, the system brightness does not drop anymore and 
the disk enters the hot steady state.  
It is interesting to note that the system approaches 
to the steady state by increasing outburst interval but 
not decreasing outburst amplitude. 
In fact, the outburst amplitude for the disk component 
remains large even in an outburst just before entering 
the steady state (see the green line in Figure 
\ref{MTBmodel-increase}). 
We see that $\dot{M}_{\rm crit}$ is around 
10$^{17.4}$~g~s$^{-1}$.  
The $V$-band magnitude in quiescence gradually 
increases with the increase in mass transfer rates, 
which is mainly due to the increasing luminosity of 
the bright spot.  

\begin{figure}[htb]
\begin{center}
\FigureFile(80mm, 50mm){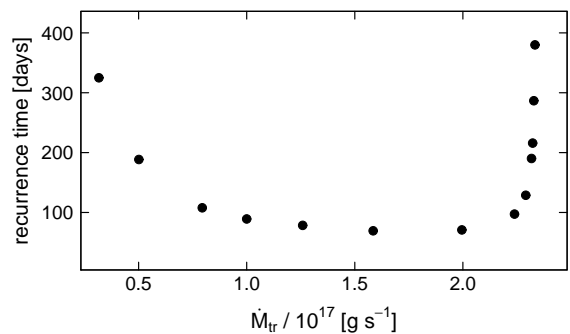}
\end{center}
\caption{
Variations of the cycle length of long outbursts against $\dot{M}_{\rm tr}$.  
}
\label{Mdot-cycle}
\end{figure}

Figure \ref{Mdot-cycle} shows the relation between the cycle 
length of long outbursts and $\dot{M}_{\rm tr}$.  
Here the cycle length is defined by the interval between 
two consecutive long outbursts because the system repeats 
one cycle with this timescale for a given $\dot{M}_{\rm tr}$ 
(see, Figure \ref{basic}). 
We see that the system approaches a steady state 
as the cycle length of long outbursts increases in 
our simulations.  
The duration of long outbursts becomes infinity 
if the mass transfer rate approaches $10^{17.4}$~g~s$^{-1}$.  
We note here that the cycle length abruptly increases and 
approaches very rapidly to infinity in figure \ref{Mdot-cycle}.
The transition from the dwarf nova outbursting stage 
to the nova-like stage with a steady disk occurs within 
a very narrow range in the mass transfer rate: a situation 
for a favorable condition for the Z Cam phenomenon.

We may note that similar behavior was also found between 
the supercycle length and the mass transfer rate 
in the thermal-tidal instability model for the SU UMa-type 
stars by \citet{osa95eruma}, who tried to explain ER UMa stars: 
DNe undergoing frequent superoutbursts 
\citep{kat95eruma,rob95eruma}.  
In this model, when the mass transfer rate is increased 
in numerical simulations, the recurrence time decreases 
at first and reaches a minimum value at a certain point 
but it begins to increase after that because the duration 
of superoutbursts increases.  
The cycle length eventually becomes infinity, and the system 
becomes a nova-like star.

\begin{figure}[htb]
\begin{center}
\FigureFile(80mm, 50mm){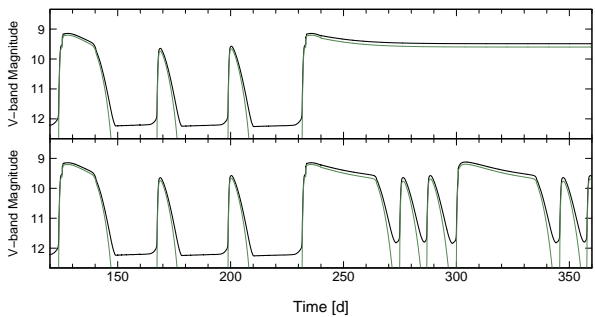}
\end{center}
\caption{
Time evolution of the $V$-band magnitude of the system for the enhanced mass transfer burst model.  
In the upper and lower panels, the transfer rates are raised up to 10$^{17.4}$~g~s$^{-1}$ and 10$^{17.3}$~g~s$^{-1}$ after 240~d, respectively.  
}
\label{MTBmodel-vband}
\end{figure}

By taking into account the above-mentioned results, 
we next try to simulate the 2021 anomalous event 
in SS Cyg based on the enhanced mass transfer burst model.  
We perform two simulations by switching the mass 
transfer rate from the standard value of 
$\dot{M}_{\rm tr}=10^{16.9}$~g~s$^{-1}$ to some enhanced 
ones after 240~d.  
The two enhanced mass transfer rates adopted here are 
10$^{17.4}$~g~s$^{-1}$ and 10$^{17.3}$~g~s$^{-1}$.  
We mark these mass transfer rates by the numbers 1 and 2 
in Figure \ref{MTBmodel-increase}.  

The results are shown in Figure \ref{MTBmodel-vband} for 
the $V$-band light curves of these two simulations.  
We see that the system enters into an (almost) 
constant-luminosity state, such as a standstill in the case of 
the highest mass transfer shown in the upper panel 
in Figure \ref{MTBmodel-vband}. 
However, the simulated light curve did not look like 
the observed 2021 anomalous event of SS Cyg, 
as the simulated constant light level was as high as 
the maximum of the short outburst  before the switching 
in mass transfer rates and it was much higher than 
the observed level averaged in the 2021 anomalous event 
of SS Cyg.  
In the second case, shown in the lower panel of 
Figure \ref{MTBmodel-vband},     
the outbursts still occur after enhanced mass transfer.  
The outburst amplitude in simulations was, however, $\sim$3~mag, 
which is much higher than the observed one ($\sim$1~mag) 
for the 2021 anomalous event in SS Cyg.  Furthermore, 
long outbursts inevitably occur in such a case 
because of enhanced mass transfer.   
The simulated light curves do not match well with 
the observed one for the 2021 anomalous event in SS Cyg.  
We thus conclude that the enhanced mass transfer is 
very unlikely to be a cause for this event.

\subsection{Enhanced viscosity model}

In order to reproduce the clear-cut outburst and quiescence 
in DNe by the thermal viscous instability model, 
it is known that we need to choose $\alpha_{\rm hot}$ 
in the hot state higher by a factor of 3--10 
than $\alpha_{\rm cool}$ for the cool state 
\citep{mey84ADtransitionwave,osa89suuma}.  
It is widely accepted that the viscosity in the hot ionized 
disk is produced by the magneto-rotational instability (MRI; 
\cite{BalbusHawley}).  
On the other hand, the origin of viscosity in the cold disk 
is not known yet.  
In the cold disk, the gas is mostly neutral, and 
the magnetic field may decouple from the disk matter, and 
the MRI turbulence could not be maintained in such a case.  
Under such a circumstance, several different possible sources 
for the viscosity in the cold disk are discussed, but so far, 
no model is widely accepted.  
It is a common practice that light curve simulations are 
performed by choosing the viscosity parameter $\alpha_{\rm cool}$ 
for the cold disk in such a way to reproduce observed light 
curves of DNe.  

\begin{figure*}[htb]
\begin{center}
\FigureFile(160mm, 50mm){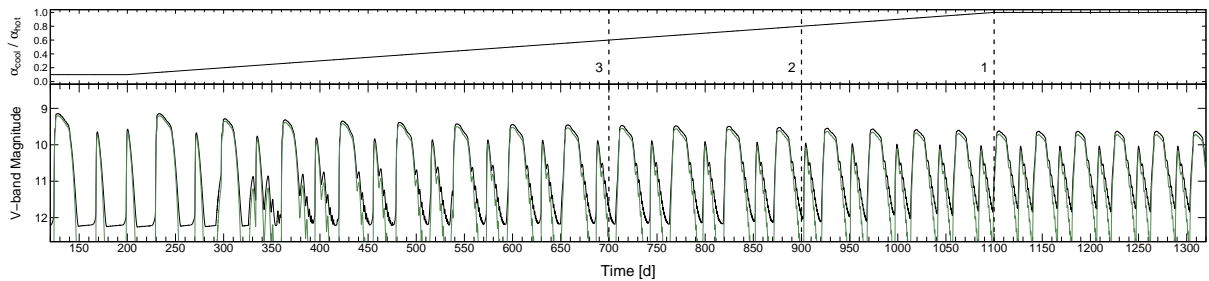}
\end{center}
\caption{
Results of our simulations in which $\alpha_{\rm cool}$
is gradually increased with time and finally becomes comparable with $\alpha_{\rm hot}$.  
(Upper) the ratio of viscosity parameters $\alpha_{\rm cool} / \alpha_{\rm hot}$. 
(Lower) the light curve for the $V$-band magnitude.
See the caption of Figure \ref{basic} as for the black and green lines in the lower panel.
}
\label{EVmodel-increase}
\end{figure*}

We here investigate the possibility of enhanced viscosity 
in the cold disk for the 2021 anomalous event in SS Cyg. 
We firstly perform simulations by gradually increasing 
the ratio of viscosity parameter $\alpha_{\rm cool} / 
\alpha_{\rm hot}$ from the standard value of 0.1 to 1.0 
to see what happens.  
Figure \ref{EVmodel-increase} shows 
the result of our simulations, where $\alpha_{\rm cool}$ 
is gradually increased during 900~d from the standard value 
from date 200~d to $\alpha_{\rm hot}$.
We see in Figure \ref{EVmodel-increase} 
that slow-rise outbursts (inside-out outbursts) more frequently 
occur as $\alpha_{\rm cool}$ increases.  
Small reflares emerge on the fading tail of outbursts.  
We also find that the flat-bottomed 
quiescence disappears.
In our simulations, we have chosen the inner edge of the disk 
near the WD surface.  
However, the inner disk in quiescence is, in reality, most 
likely truncated by the evaporation of the disk gas 
via coronal siphon flow \citep{mey94siphonflow} to produce 
the hot optically thin corona near the WD, which produces 
X-ray radiation.  
The small-amplitude inside-out outbursts are mostly likely 
suppressed, and the quiescent duration may be longer 
in such a situation.  
Also, small reflares on the fading tail of outbursts 
may disappear by taking into account the variation of 
$r_{\rm in}$ \citep{men00BHXN}.

As the viscosity in the cold disk is increased further 
in Figure \ref{EVmodel-increase}, the outburst cycle becomes 
shorter, and the quiescent brightness level increases.  
Also, the difference in the amplitude and duration between long 
and short outbursts become smaller.  
This behavior looks similar in some sense to the observed 
light curve before the 2021 anomalous event in SS Cyg.  
As discussed in \citet{kim21sscyg}, low-amplitude and 
slow-rise outbursts seem to have frequently occurred after 
around BJD 2458700, the long outbursts were suppressed, 
and the cycle length of outbursts became shorter 
(see Figure \ref{lc-normal}).  

\begin{figure}[htb]
\begin{center}
\FigureFile(80mm, 50mm){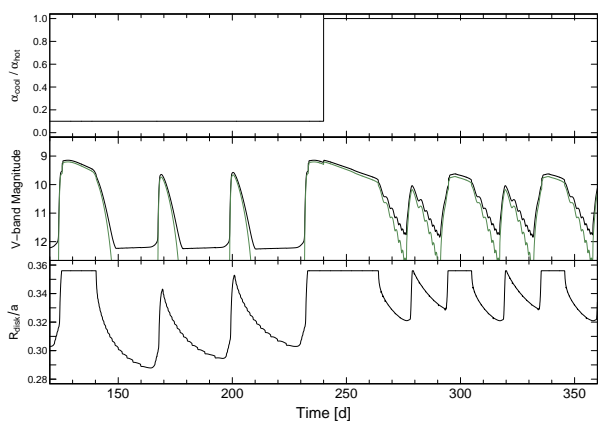}
\end{center}
\caption{
Results of our simulations for the enhanced viscosity model.  Here, $\alpha_{\rm cool} / \alpha_{\rm hot}$ jumps from 0.1 to 1.0 on 240~d.  
From top to bottom: the ratio of viscosity parameters $\alpha_{\rm cool} / \alpha_{\rm hot}$, the light curve for the $V$-band magnitude, and the disk radius in units of the binary separation. 
See the caption of Figure \ref{basic} as for the black and green lines in the middle panel.  
}
\label{EVmodel}
\end{figure}

\begin{figure}[htb]
\begin{center}
\FigureFile(80mm, 50mm){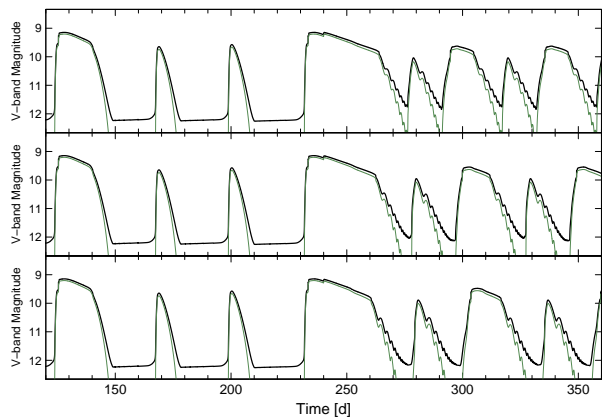}
\end{center}
\caption{
Time evolution of the $V$-band magnitude of the system for the enhanced viscosity model.  Here, $\alpha_{\rm cool}$ is enhanced on 240~d.
The top panel is the same as the middle panel of Figure \ref{EVmodel}.
In the middle panel, $\alpha_{\rm cool} / \alpha_{\rm hot}$ is raised up to 0.8.  
In the bottom panel, $\alpha_{\rm cool} / \alpha_{\rm hot}$ is raised up to 0.6.  
}
\label{EVmodel-vband}
\end{figure}

We understand the effect on the light curve due to 
enhanced viscosity from the change in the `S'-shaped 
thermal equilibrium curve.  
If the viscosity in the cool state is enhanced, 
the unstable branch in the thermal equilibrium curve 
becomes shorter (see the right panel of Figure \ref{scurves}).
Then the upper limit of the effective temperature 
in the quiescent state increases.
This change also prevents cooling and heating waves 
from propagating over the entire disk and causes 
the thermal instability frequently in local regions.  
The innermost region of the disk stays for a longer time 
in the hot state, and the thermal instability 
frequently occurs in other regions.
Besides, more mass is conveyed inwards, and the thermal 
instability is easily triggered in regions other than 
the outermost disk.  
This is why we confirm small-amplitude and slow-rise 
outbursts and small reflares in the fading tail 
in the simulated light curves.  

Let us now try to simulate the 2021 anomalous event 
in SS Cyg by artificially increasing the value of 
$\alpha_{\rm cool} / \alpha_{\rm hot}$. 
Figure \ref{EVmodel} illustrates the results of our simulations 
in which $\alpha_{\rm cool} / \alpha_{\rm hot}$ is increased 
from 0.1 to 1.0 instantaneously on 240~d.  
We mark this switching ratio as the number 1 in 
Figure \ref{EVmodel-increase}.  
We see in Figure \ref{EVmodel} that the minimum level 
in the simulated light curve increases around 0.5 mag 
after the switching in this ratio.  
The saw-tooth-like oscillatory light curves (or 
outbursts with a rapid-rise and slow-decay ones) 
result with an amplitude of 2 mag.  
Qualitatively speaking, the model light curve after 
the switching event in Figure \ref{EVmodel-increase} 
looks like the 2021 long outburst and its subsequent 
anomalous phenomenon in SS Cyg: three saw-tooth-like 
outbursts have occurred within about 60 days after 
the local minimum at 275~d.
However, quantitatively speaking, the oscillatory amplitude 
of about 2 mag in our simulations is larger than 
the observed one with about 1 mag in the 2021 anomalous event, 
and the increment in the minimum brightness in our simulations 
is smaller than the observed one of about 1.5 mag.

We have tried two more simulations basically in the same way 
as that mentioned above but with different increments 
in $\alpha_{\rm cool}$ and we compare them in Figure 
\ref{EVmodel-vband}.  
In the middle panel, $\alpha_{\rm cool} / \alpha_{\rm hot}$ 
is increased to 0.8, while it is increased to 0.6 
in the bottom panel.  
The final $\alpha_{\rm cool} / \alpha_{\rm hot}$ values in 
these two simulations are denoted by the numbers 2 and 3 
in Figure \ref{EVmodel-increase}.  
We see that the larger the increment is, the smaller the amplitude of 
outbursts after the long outburst starting from $\sim$240~d 
is and shorter the recurrence time of outbursts is.  
We have tried the same simulations with different $r_{\rm in}$ 
and $\alpha$ values and confirm that the main features, 
as mentioned above, are always reproduced.  

\begin{figure*}[htb]
\begin{center}
\vspace{-5mm}
\FigureFile(160mm, 50mm){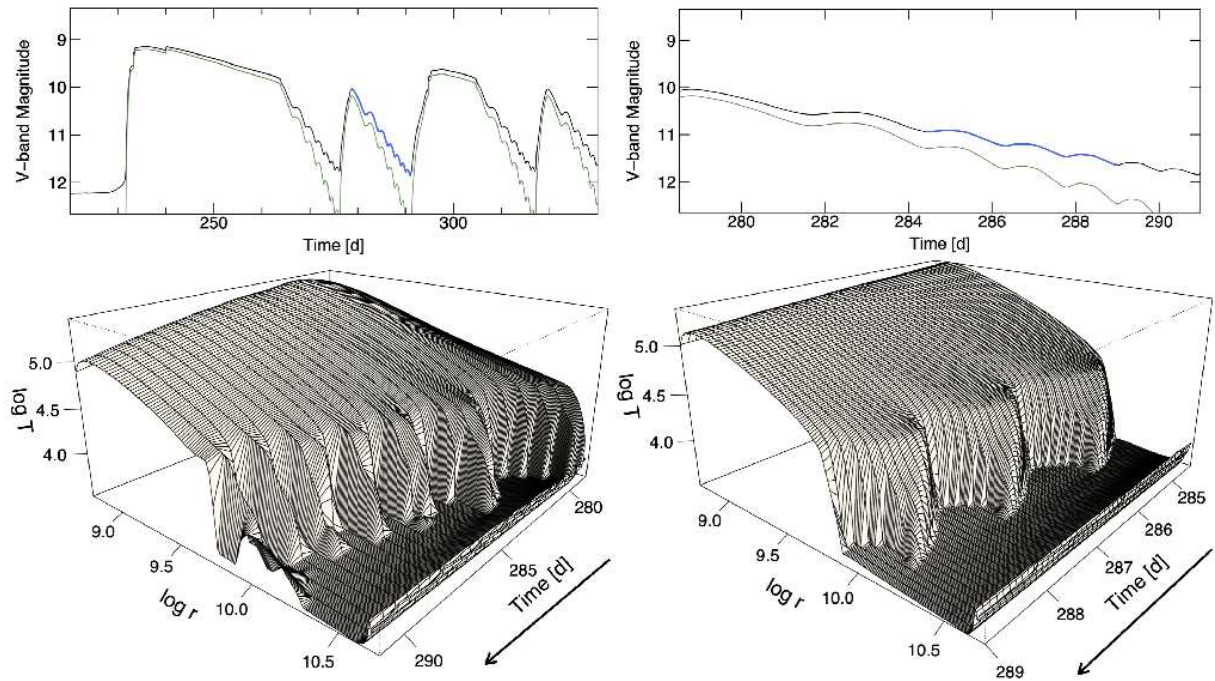}
\end{center}
\caption{
Three-dimensional figures of cooling front propagation (lower panels) and corresponding light curves (upper panels) of the simulation displayed in Figure \ref{EVmodel}.  
(Left) The time evolution of the disk temperature from 278.5~d to 291~d per 0.5~d.  
The corresponding time period in the lower panel is represented by the blue line in the upper light curve.  
(Right) The detailed version of a part of cooling front propagation.  The time evolution of the disk temperature from 284.4~d to 289~d per 0.1~d and the corresponding light curve.  
}
\label{front-propagation}
\end{figure*}

\begin{figure*}[htb]
\begin{center}
\vspace{-5mm}
\FigureFile(160mm, 50mm){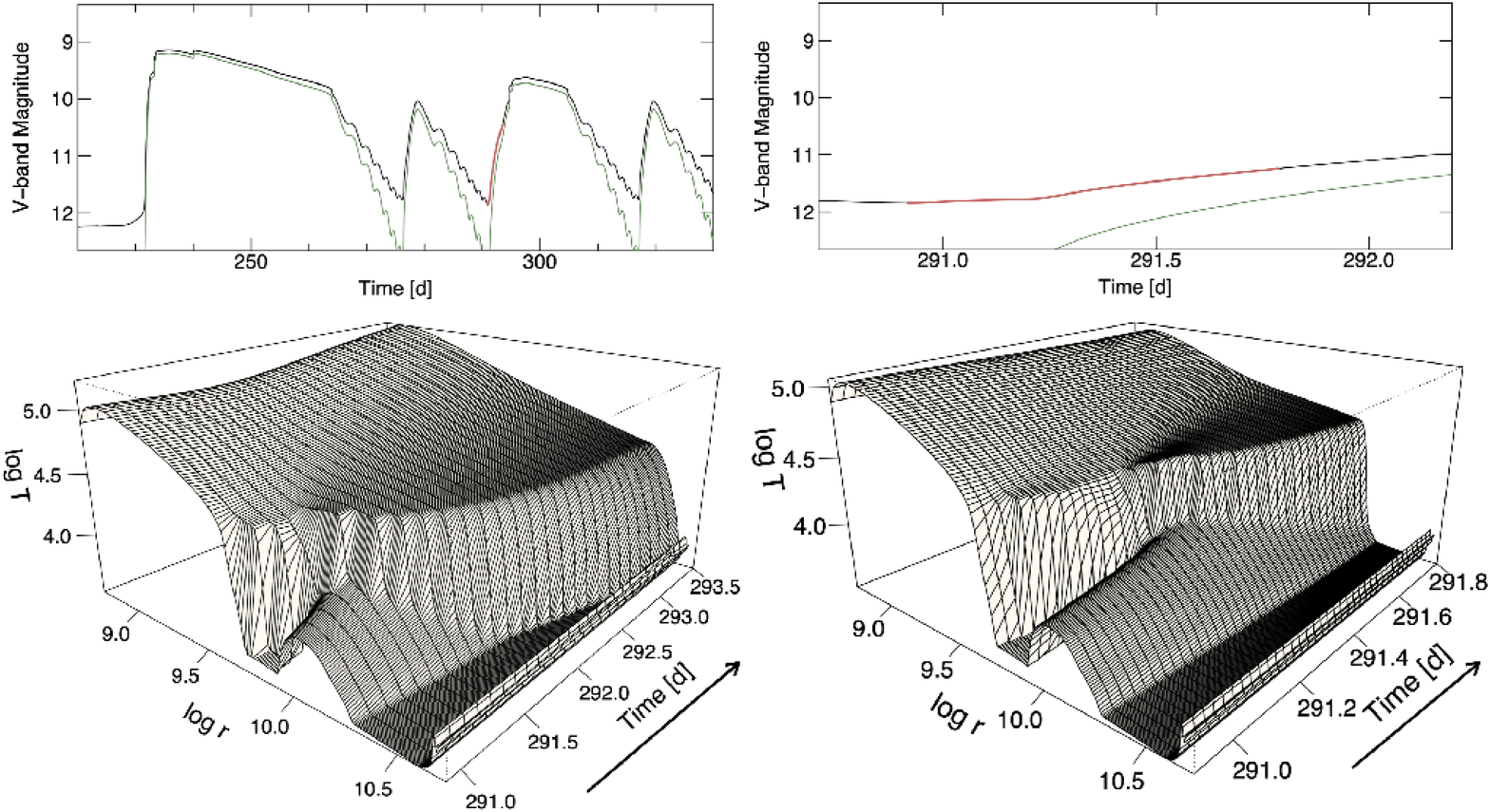}
\end{center}
\caption{
Three-dimensional figures of heating front propagation (lower panels) and corresponding light curves (upper panels) of the simulation displayed in Figure \ref{EVmodel}.  
(Left) The time evolution of the disk temperature from 290.9~d to 293.5~d per 0.1~d.  The corresponding time period in the lower panel is represented by the red line in the upper light curve.  
(Right) The detailed version of a part of heating front propagation.  The time evolution of the disk temperature from 290.7~d to 292.2~d per 0.025~d and the corresponding light curve.  
}
\label{front-propagation-2}
\end{figure*}

Before we tried our simulations, we had anticipated that 
the simulated light curves would result in small-amplitude 
fluctuations in the case of constant $\alpha$ for 
the hot and cool states.
However, our results shown in Figure \ref{EVmodel} demonstrate 
that fairly clear outbursts occur rather than small-amplitude 
fluctuations even in the case of 
$\alpha_{\rm cool} / \alpha_{\rm hot}$ = 1.0 
(in particular, see the green line in Figure \ref{EVmodel} 
for the disk component only).  
For instance, \citet{min85DNDI} showed that the simulations 
with $\alpha_{\rm cool} / \alpha_{\rm hot}$ = 1.0 results in  
small-amplitude fluctuations.
Also, \citet{mey84ADtransitionwave} and \citet{min90irradiation} 
explored the conditions for clear-cut outbursts in which 
the heating and cooling fronts propagate all the way to the disk 
from the outer edge to the inner edge.  
One of such conditions is that $\Sigma_{\rm max} / \Sigma_{\rm min}$ 
is larger than 2, where $\Sigma_{\rm min}$ is the critical 
surface density at the downward transition from the hot state. 
The reason is that if $\Sigma_{\rm max} / \Sigma_{\rm min}$ 
is less than 2, the heating and cooling fronts cannot travel 
a long distance as they are reflected within a short distance.
As seen in the S-shaped thermal-equilibrium curves shown 
in the left panel of Figure \ref{scurves}, $\Sigma_{\rm max} 
/ \Sigma_{\rm min}$ is slightly less than 2 in our case of 
$\alpha_{\rm cool} / \alpha_{\rm hot}$ = 1.0 
and thus, our cases do not satisfy this condition.

To see what happens in our simulations in the case of 
constant $\alpha$ shown in Figure \ref{EVmodel}, 
we show figures for the time evolution of the transition front.
Figure \ref{front-propagation} illustrates three dimensional 
plots of the propagation of a cooling wave in 
a part of simulation results with $\alpha_{\rm cool} / 
\alpha_{\rm hot}$ = 1.0, which are displayed in 
Figure \ref{EVmodel}. 
The three-dimensional plots for 
the temperature distribution in the lower panels show 
a part of the light curves indicated by the blue line 
in the upper panels, and the right-hand figures correspond 
to a locally expanded version of the left-hand panel 
with further details.
We see from the left panel of Figure \ref{front-propagation} 
that the cooling front propagates inward and reaches
deep inner disk to $r < 10^{10}$~cm. 
However, its propagation is not smooth, but it goes locally 
back and forth, which is seen as a small ripple-like 
variation in the decay light curve. 
The right panel of Figure \ref{front-propagation} is 
its locally expanded version, and it shows very clearly 
this feature. 
However, the cooling front never reaches the innermost 
region of the disk, but it is reflected in the middle of 
the disk as a heating front. 
The innermost part of the disk thus remains always 
in the hot state in our simulation. 

\textcolor{black}{
The reason why our simulation results in a clear outburst 
with amplitudes as large as 2 mag in the case of 
constant $\alpha$ is not certain, but the results 
in the case of constant $\alpha$ may be very sensitive to 
the detailed profile of the S-shaped thermal-equilibrium 
curve used in the simulations. 
The value of $\Sigma_{\rm max} / \Sigma_{\rm min}$ 
in our case is less than 2 but very close to 2.  
Besides that, the maximum effective temperature of the cool 
branch is as low as 4,000~K in the outer disk in our case 
(see the left panel of Figure \ref{scurves}). 
The separation of the effective temperature between 
the hot branch and the cool branch in the thermal equilibrium 
curve in our case is thus wider than those in other works, 
as described in section 2. 
These two effects combined could be the reason for this. 
We here note that a clear outburst found in our case is not 
a genuine full-scale outburst because the innermost part of 
the disk always remains in the hot state. 
The problem is then how far (how deeply) the cooling front 
propagates in the disk, and this may depend on 
the thermal equilibrium curve used, so the amplitude of 
oscillatory light variations may be different with 
a different thermal-equilibrium curve.
This cannot, however, be easily tested because different 
authors use different thermal-equilibrium curves.
}

Figure \ref{front-propagation-2} illustrates three 
dimensional plots of the propagation of a heating wave, 
a corresponding one to Figure \ref{front-propagation} 
for a cooling wave.  
As in Figure \ref{front-propagation}, 
the three-dimensional plots for the temperature 
distribution in the lower panels correspond to 
a part of the light curves indicated by the red line 
in the upper panels, and the right-hand figures are 
a locally expanded version of the left-hand figures.
We see from Figure \ref{front-propagation-2} 
that the heating wave, which is generated by the reflection 
of the cooling wave in the middle of the disk, propagates 
outward (i.e., the inside-out type) and its propagation 
is rather smooth in this case. 
Its right panel is a locally expanded version of the left panel 
and it clearly shows no ripple-like structure in contrast 
with the case of the cooling front. 
It was found that the heating wave propagated faster 
than the cooling wave, in this case, resulting in a rapid-rise 
and a slow-decay light curve. 
It is generally believed that the rise and decay of light curves 
are symmetric in the inside-out type outbursts.
The reason why we have obtained a rapid-rise and slow-decay 
type light curve in our case may probably be because 
the heating front propagated smoothly while the cooling front 
did not so.  

\begin{figure*}[htb]
\begin{center}
\FigureFile(160mm, 50mm){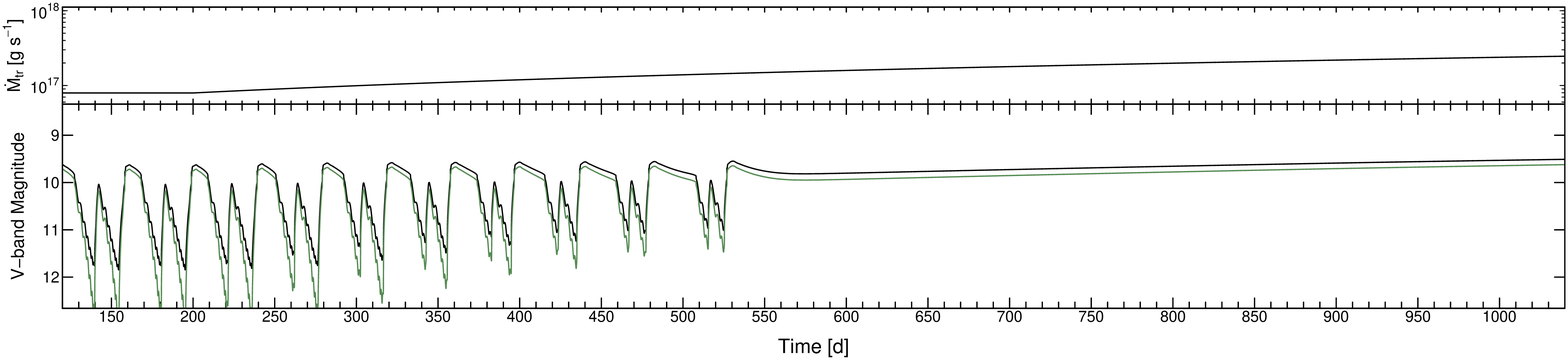}
\end{center}
\caption{
Results of our simulations in which the mass transfer rate gradually increases with time in the case of $\alpha_{\rm cool} / \alpha_{\rm hot} = 1.0$.  
See the caption of Figure \ref{basic} as for the black and green lines in the lower panel.  
}
\label{MTBmodel-increase-2}
\end{figure*}

In the previous subsection, we have demonstrated 
in the case of $\alpha_{\rm cool} / \alpha_{\rm hot}$ = 0.1 
that the system approaches a hot steady state by 
increasing the duration of the long outburst 
(see Figures \ref{MTBmodel-increase} and \ref{Mdot-cycle}) 
but not by decreasing the outburst amplitude, 
when we increase the mass transfer rate from that of 
the standard case up to the critical one.
As described above, the outburst amplitude becomes 
smaller when the viscosity in the cool state is enhanced.  
We wonder whether or not there is another possibility 
(i.e., the second possibility) that the system might 
approach to a steady state by decreasing the amplitude of 
outburst and finally with vanishing outburst amplitude 
in the case of  $\alpha_{\rm cool} / \alpha_{\rm hot} 
\simeq$ 1.0, when we increase mass transfer rate with time. 
 Figure \ref{MTBmodel-increase-2} illustrates our simulations 
in the case of $\alpha_{\rm cool} / \alpha_{\rm hot} = 1.0$ 
in which the mass transfer rate is increased with time.  
We see from this figure that the amplitude of outbursts 
becomes smaller with the increase in mass transfer rates 
but never vanishingly small when the system approaches to 
a hot steady state. 
By comparing this case with the case where $\alpha_{\rm cool} / 
\alpha_{\rm hot} = 0.1$, the outburst amplitude is smaller 
by $\sim$1.0 mag just before the system enters into 
a constant luminosity state in this case.  
Besides, $\dot{M}_{\rm crit}$ is lower in this case 
since the disk easily becomes thermally stable with higher 
ratios of $\alpha_{\rm cool} / \alpha_{\rm hot}$.  
However, the approach to a hot steady state is basically 
the same between the two cases, that is, the duration of 
long outburst becomes infinity.  
Other simulations we have performed give basically 
the same result.  The second possibility seems to be ruled 
out in our formulation, where the tidal truncation 
and variable disk radius are taken into account.

\subsection{Possibility of gas-stream overflow}

The most important observational feature of the 2021 
anomalous event in SS Cyg is an elevation of brightness level 
in quiescence, that is, from 12 mag in the ordinary quiescence 
to $10 \sim 11$~mag during the anomalous event.  
If the entire disk stays in the cool state with low viscosity 
in quiescence, such an anomalous event is obviously not possible.  
We need to have a hot state at least in a part of the disk, 
in particular in the inner part of the disk.  
In this respect, gas stream overflow could be another possible 
solution to this problem.  
In fact, \citet{kim20tiltdiskmodel} performed light curve 
simulations in which the effect of gas stream overflow plays 
an important role in the case of tilted disks in relation with 
the IW And-type phenomenon. 
Even in the case of a non-tilted disk, a part of the gas stream 
from the secondary star can overflow above and below the disk edge 
to penetrate into the inner part of the disk, as discussed firstly 
by \citet{lub76stream}.  
\citet{hes99streamoutflow} showed that the gas stream could overflow 
the disk edge, in particular in the case of the cold disk in quiescence. 
\citet{kun01streamdiskoverflow} performed the smoothed particle 
hydrodynamic simulations on the steam overflow, showing that 
a significant fraction of the gas stream can overflow 
the disk edge and penetrate into the inner disk.  

As for the light curve simulations for DN outbursts based on 
the thermal-viscous instability model, \citet{sch98streamoverflow}
studied the effects of the stream overflow upon the outburst 
light curves.  
However, their study was rather limited for a special case.  
In this study, we perform light curve simulations to explore 
the effects of the stream overflow for much wider ranges 
in parameter space.  

\begin{figure}[htb]
\begin{center}
\FigureFile(80mm, 50mm){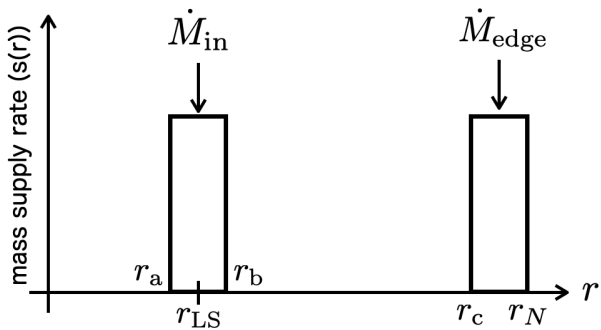}
\end{center}
\caption{
Mass supply pattern for the gas-stream overflow.  The mass transfer rate is divided in $\dot{M}_{\rm in}$ and $\dot{M}_{\rm edge}$ in this case.  
}
\label{stream-overflow-fig}
\end{figure}

To treat the gas-stream overflow, we change the mass supply 
pattern to the disk.  
If the gas stream overflows the outer disk rim, a part of 
the stream enters the outer disk edge, and the rest of it 
flows into the inner disk.  
In this study, we assume that the overflowing 
gas entering into the inner disk basically goes to 
a region around the Lubow-Shu radius (i.e., the 
circularization radius). 
There are two reasons why we adopt this assumption.
Firstly, the hydrodynamic simulations by 
\citet{kun01streamdiskoverflow} 
demonstrate that the overflowing gas stream settles 
at a region around the circularization radius, thus 
justifying our assumption. 
The second reason concerns a problem of numerical 
difficulty which \citet{kim20tiltdiskmodel} 
encountered in the case when the gas stream reaches 
the region of $r < r_{\rm LS}$ in the cold disk. 
Its details were discussed in section 6.3 of 
\citet{kim20tiltdiskmodel}. 
In short, the disk matter inside of $r_{\rm LS}$, 
to which the overflowing gas is added, tends to move 
towards $r_{\rm LS}$ in the cold disk, since 
the transferred gas has the specific angular momentum 
of $\sqrt{G M_1 r_{\rm LS}}$. 
This causes thinning of the disk matter there, which 
results in a stoppage of calculations.  
To avoid this numerical difficulty, we assume in our 
formulation that the mass of overflowing gas is 
exclusively supplied to a region around $r_{\rm LS}$ 
as a delta-function-like form.
We therefore set the mass supply rate $s(r)$ as displayed in 
Figure \ref{stream-overflow-fig}. 
Here, $\dot{M}_{\rm edge}$ and $\dot{M}_{\rm in}$ are 
the transferred mass to the outer disk rim and 
that to the region near $r_{\rm LS}$, respectively.  
We define the ratio of $\dot{M}_{\rm in}$ to the total mass 
transfer rate as $f_{\rm overflow}$ and 
$\dot{M}_{\rm edge} = (1 - f_{\rm overflow}) \dot{M}_{\rm tr}$ and 
$\dot{M}_{\rm in} = f_{\rm overflow} \dot{M}_{\rm tr}$.  
In this formulation, $s(r)$ is expressed as follows: 
\begin{eqnarray}
s(r) &=& \frac{\dot{M}_{\rm edge}}{0.02a}~(r_{\rm c} \leq r \leq r_N), \\
s(r) &=& \frac{\dot{M}_{\rm in}}{0.02a}~(r_{\rm a} \leq r \leq r_{\rm b}),
\label{source-term-overflow}
\end{eqnarray}
where $r_{\rm a}$ is the border between the two meshes which is 
closest to $r_{\rm LS} - 0.01a$, $r_{\rm b}$ is that closest to 
$r_{\rm LS} + 0.01a$, $r_{\rm c}$ is that closest to $r_{N} - 0.02a$, 
respectively.  

\begin{figure*}[htb]
\begin{center}
\FigureFile(160mm, 50mm){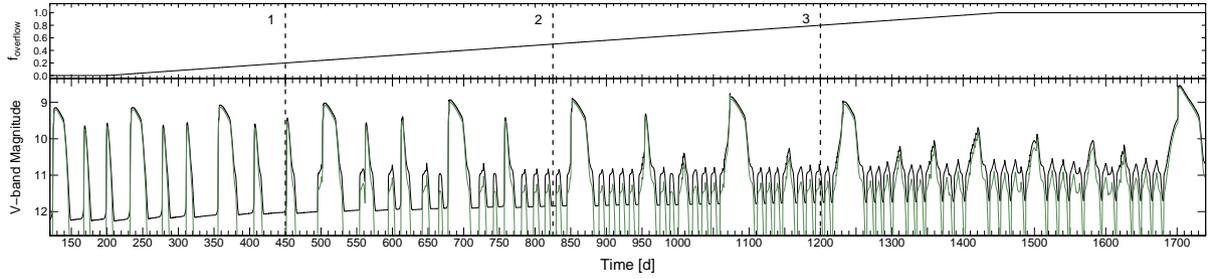}
\end{center}
\caption{
Our results of simulations in the case of gas-stream overflow in which $f_{\rm overflow}$ is gradually increased with time.  The lower panel shows the $V$-band light curve and the upper panel does the corresponding $f_{\rm overflow}$ value.
See the caption of Figure \ref{basic} as for the black and green lines in the lower panel.  
}
\label{overflow-increase}
\end{figure*}

We have carried out simulations in the case of the gas-stream 
overflow in which we gradually increase $f_{\rm overflow}$ 
with time while the other parameters are kept to be 
the same as those of the standard model.  
Figure \ref{overflow-increase} illustrates our simulation results.  
Here, we calculate the $V$-band magnitude of the bright spot 
by assuming that its luminosity is a sum of 
$0.25 G M_1 \dot{M}_{\rm edge} / r_N$ and 
$0.25 G M_1 \dot{M}_{\rm in} / r_{\rm LS}$.  
The size of each spot is postulated to be 2\% of the disk size 
as in \citet{kim20tiltdiskmodel}.  
The temperature of the bright spot is determined by 
$T_{\rm BS} = (L_{\rm BS} / (\sigma \delta S))^{1/4}$, 
where $L_{\rm BS}$ is the luminosity of the bright spot, 
$\sigma$ is the Stefan–Boltzmann constant, and 
$\delta S$ is the size of each bright spot. 
Although it depends on $\dot{M}_{\rm edge}$ and $\dot{M}_{\rm in}$, 
it is typically $\sim$10,000~K.
As $f_{\rm overflow}$ increases, the cycle length of long outbursts 
becomes longer and a new type of outburst with a small-amplitude 
about 1~mag begins to appear beside the ordinary short and 
long outbursts and its number is increasing with the increase 
in $f_{\rm overflow}$. 
All of the outbursts become inside-out type, and the quiescent 
level increases with the increase in $f_{\rm overflow}$ 
because of an increase in brightness of the inner bright spot.  
When $f_{\rm overflow}$ becomes greater than 0.8, the inner part of 
the disk stays in the hot state most of the time. 
In the special case of $f_{\rm overflow} = 1.0$, all of mass 
transferred from the secondary is supplied to the inner disk and 
no mass is supplied to the outer disk edge, i.e., $\dot{M}_{\rm edge}$ 
is zero. 
In such a case, the outer disk edge expands and reaches 
the tidal truncation radius even in the cool state and the mass 
slowly accumulates there.  
Eventually, the heating wave produced by the inside-out outburst 
propagates outward and reaches the outer edge of the disk, 
i.e., the tidal truncation radius, producing a long and large 
outburst, as seen as the last outburst in Figure \ref{overflow-increase}.
The outer edge of the disk remains at the tidal truncation radius 
even when the outburst has ended and the outer part of the disk 
returns to the cool state in this case.  

\begin{figure}[htb]
\begin{center}
\FigureFile(80mm, 50mm){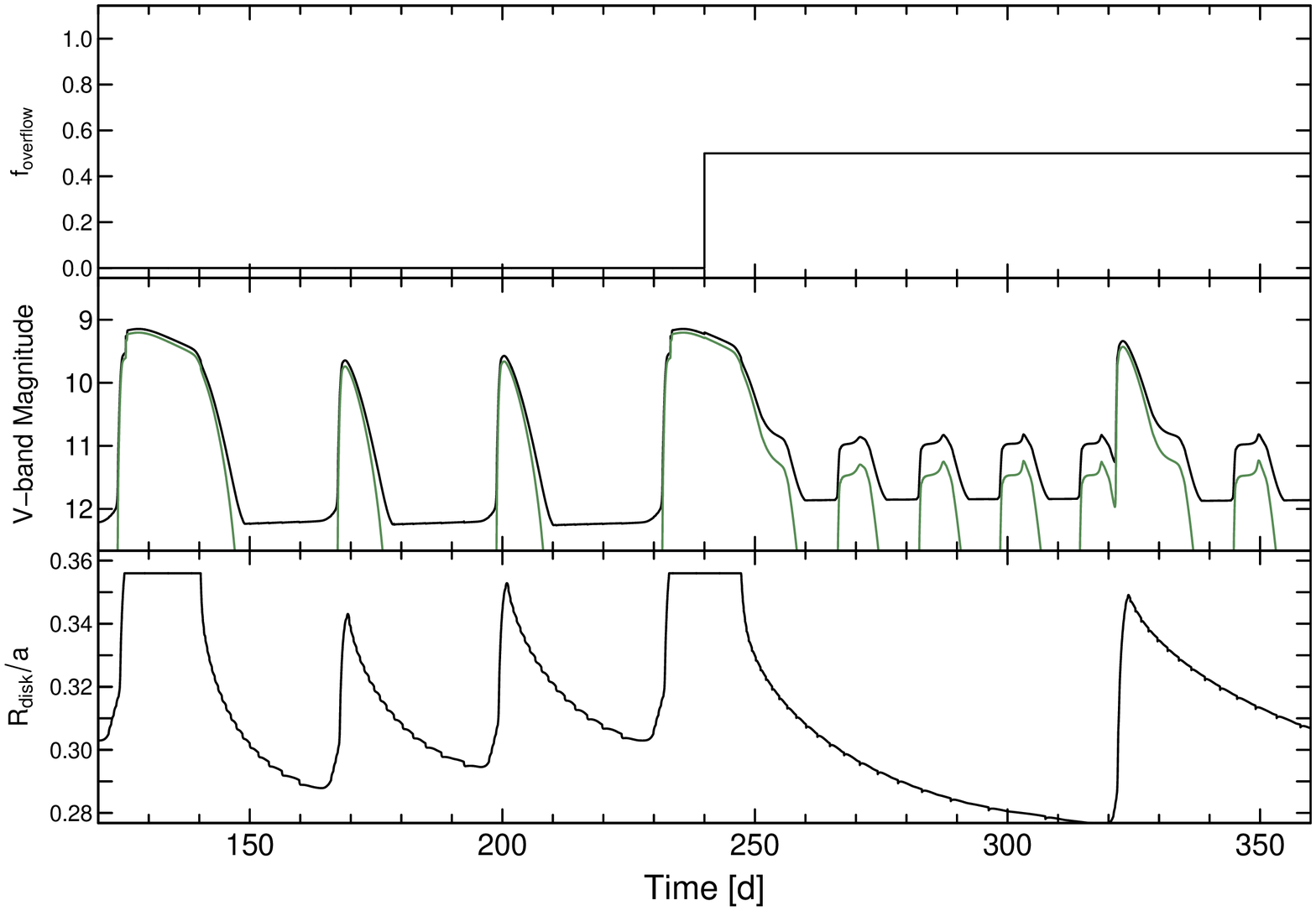}
\end{center}
\caption{
Results of our simulations in the case where $f_{\rm overflow}$ is amplified on 240~d. 
From top to bottom: variations in $f_{\rm overflow}$, the light curve for the $V$-band magnitude, and the disk radius in units of the binary separation. 
Here, $f_{\rm overflow}$ is 0.5 after 240~d.  
See the caption of Figure \ref{basic} as for the black and green lines in the middle panel.  
}
\label{overflow-model}
\end{figure}

\begin{figure}[htb]
\begin{center}
\FigureFile(80mm, 50mm){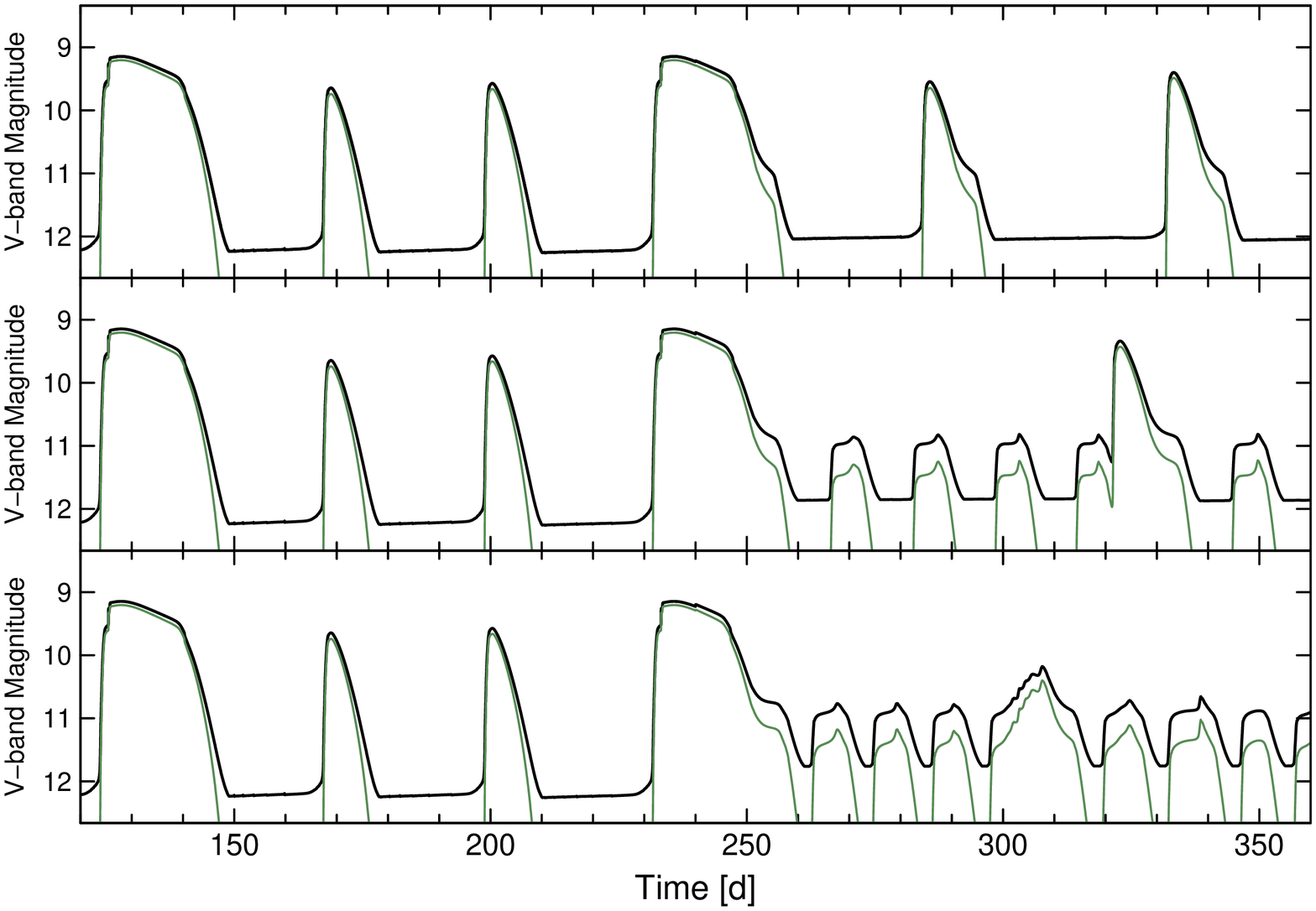}
\end{center}
\caption{
Time evolution of the $V$-band magnitude of the system for the overflow model.  
The middle panel is the same as that in the second panel of Figure \ref{overflow-model}.  
In the top and bottom panels, $f_{\rm overflow}$ is raised up to 0.2 and 0.8 after 240~d, respectively.  
See the caption of Figure \ref{basic} as for the black and green lines in the middle panel.  
}
\label{overflow-vband}
\end{figure}

Let us now examine if the effect of gas-stream overflow 
could explain the 2021 anomalous event of SS Cyg.  
We have performed simulations by artificially increasing 
$f_{\rm overflow}$ from 0.0 to 0.5 (i.e., the half of 
mass supply to the inner disk and the other half to 
the outer edge) on 240~d, and the results are displayed 
in Figure \ref{overflow-model}.  
We mark this $f_{\rm overflow}$ value as the number 2 in 
Figure \ref{overflow-increase}.  We see from 
Figure \ref{overflow-model} that the cycle length of 
outbursts becomes longer after switching $f_{\rm overflow}$  
because the transferred mass to the outer disk edge 
becomes lower.  
On the other hand, inside-out-type low-amplitude 
outbursts frequently occur since a large amount of mass 
is directly transferred to the inner disk, and 
the total duration of the quiescent state becomes thus 
shorter.  
We have done two more simulations by increasing 
$f_{\rm overflow}$ to other values on 240~d.  
Figure \ref{overflow-vband} illustrates three light curves 
and the middle panel is the same as that of 
Figure \ref{overflow-model}.
We see from the top panel of Figure \ref{overflow-vband} that     
inside-out outbursts do not occur in this case with 
low $f_{\rm overflow}$ but the decaying time from 
the outburst maximum becomes longer as compared to that 
before the switching of $f_{\rm overflow}$ value.  
On the other hand, if $f_{\rm overflow}$ is high, 
the quiescent duration becomes shorter because inside-out 
outbursts frequently occur, so that luminosity dips appear 
in the light curve (see the bottom panel of 
Figure \ref{overflow-vband}).  
Any of the three light curves shown in 
Figure \ref{overflow-vband}) do not look like the 2021 
anomalous event of SS Cyg because the minimum level of 
light curves in all three simulations remained near 12~mag 
and we must conclude that the effect of gas stream overflow 
is not enough to explain this particular event.

\begin{figure*}[htb]
\begin{center}
\FigureFile(160mm, 50mm){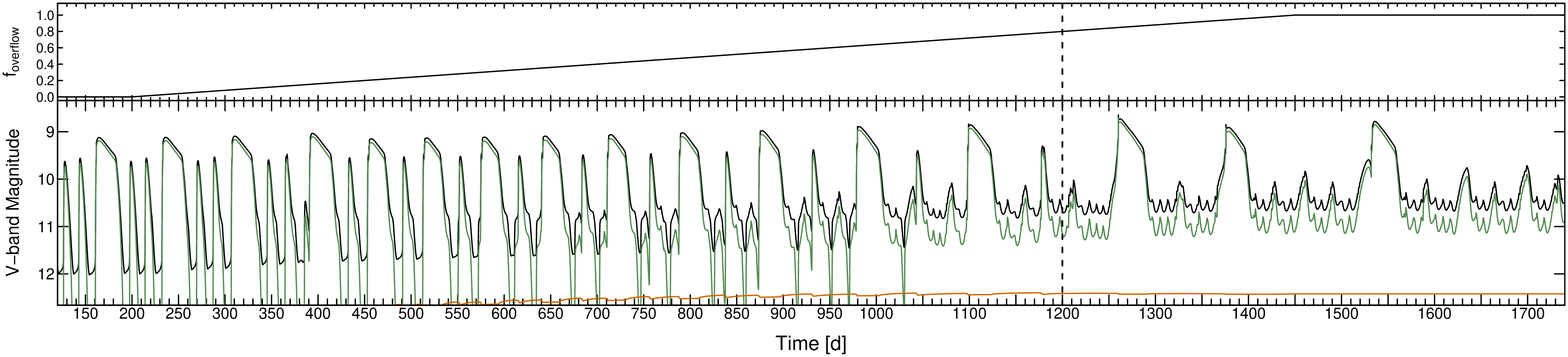}
\end{center}
\caption{
Same as Figure \ref{overflow-increase} but $\dot{M}_{\rm tr} = 1.5 \times 10^{17}$~g~s$^{-1}$.  
See the caption of Figure \ref{basic} as for the black, green, and orange lines in the lower panel.  
}
\label{overflow-increase-2}
\end{figure*}

When we had performed our simulations with the effect 
of gas-stream overflow, we have come to realize 
the similarity between our calculations and those by 
\citet{kim20tiltdiskmodel} in which the gas-stream overflow 
occurred because of a tilted disk (see, e.g., Fig.~17 of 
\citet{kim20tiltdiskmodel}). 
\citet{kim20tiltdiskmodel} simulated light curves produced 
by the thermal instability in the accretion disk tilted out 
of the binary orbital plane by implementing the mass supply 
pattern to the tilted disk under an assumption that the stream 
trajectory was calculated as the ballistic one of 
a gas particle.  Although details in the mass supply pattern 
are different between these two cases, the resultant variation 
in light curves is similar to each other where higher 
$f_{\rm overflow}$ in our case corresponds to a higher tilt 
angles there.  
By comparing two figures, Fig.~16 and Fig.~17 of 
\citet{kim20tiltdiskmodel}, we find that the inner part of 
the disk is persistently in the hot state (possibly 
a necessary condition for explaining the 2021 anomalous event 
of SS Cyg) if the mass transfer rate is a little higher than 
that of the standard model in this paper. 
To confirm this, we have performed the same type of simulations 
as Figure \ref{overflow-increase} but with a higher mass transfer 
rate than that of our standard model, and the results are shown in 
Figure \ref{overflow-increase-2}.  We see in 
Figure \ref{overflow-increase-2} that the minimum light level 
is elevated to around 10~mag if $f_{\rm overflow}$ is 
sufficiently high because the inner part of the disk is 
persistently in the hot state in such a case.  
This behavior is close to that of corresponding cases for 
tilted disks (see, Fig.~16 of of \citet{kim20tiltdiskmodel}).

\begin{figure}[htb]
\begin{center}
\FigureFile(80mm, 50mm){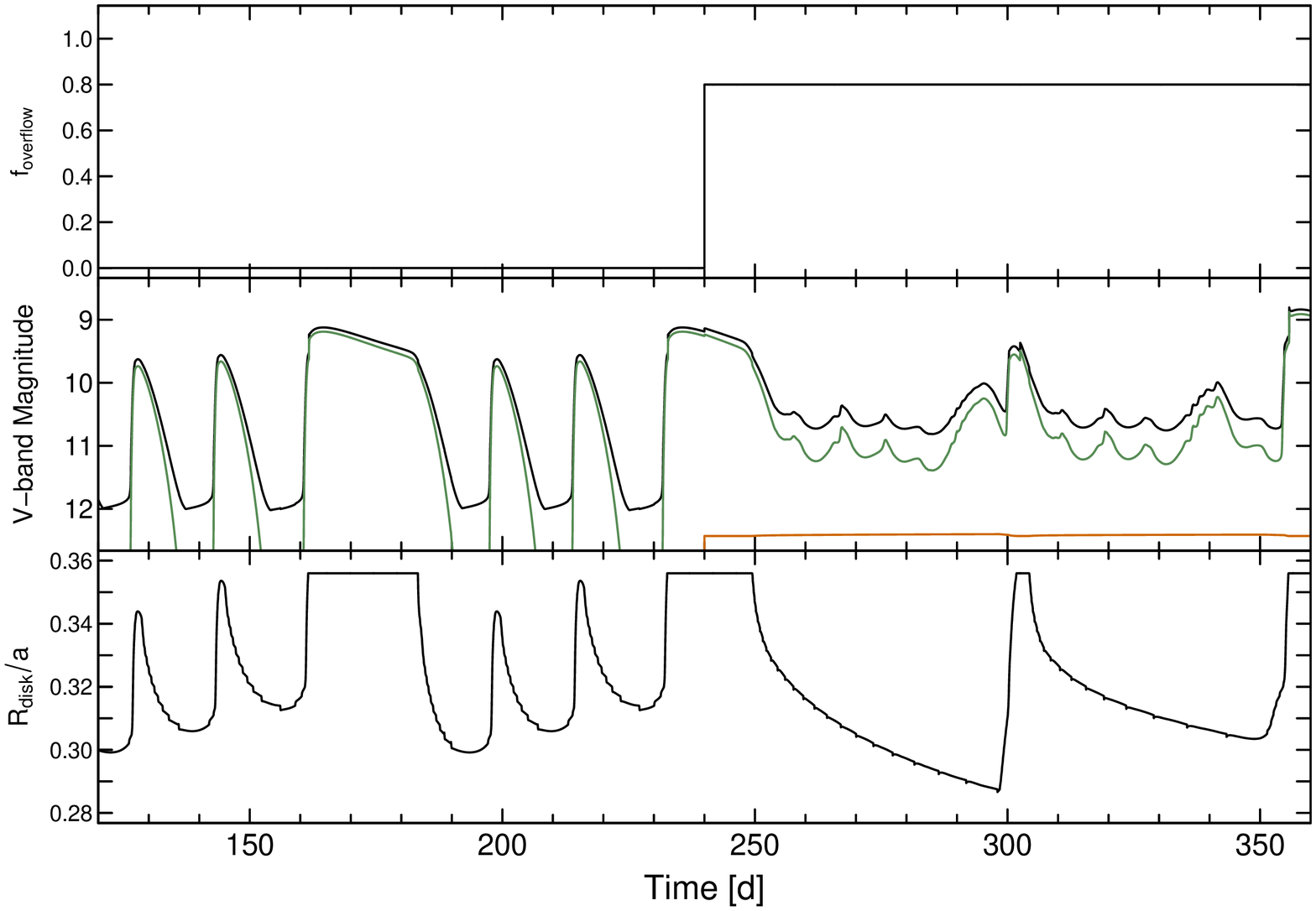}
\end{center}
\caption{
Results of our simulations in the case where $\dot{M}_{\rm tr} = 1.5 \times 10^{17}$~g~s$^{-1}$ and $f_{\rm overflow}$ changes on 240~d.  
Here, $f_{\rm overflow}$ is increased to 0.8 after 240~d.  
The corresponding $f_{\rm overflow}$ is denoted as a dashed line in Figure \ref{overflow-increase-2}.  
See the caption of Figure \ref{basic} as for the black, green, and orange lines in the third panel.  
}
\label{MTB-overflow}
\end{figure}

The mass-transfer rate in SS Cyg may be a little higher 
than $\dot{M}_{\rm tr} = 10^{16.9}$~g~s$^{-1}$, a value 
that we have adopted in the standard model (see section 2).  
We have thus tried one more simulation in which $f_{\rm overflow}$ 
is amplified to 0.8 on 240~d in the case of $\dot{M}_{\rm tr} 
= 1.5 \times 10^{17}$~g~s$^{-1}$.  
The results are displayed in Figure \ref{MTB-overflow}.  
We see from the middle panel of Figure \ref{MTB-overflow} 
that the system shows low-amplitude fluctuations 
and occasionally large brightening after 240~d.  
Also, the average flux level is as high as 10 mag 
during oscillations because the inner disk always stays 
in the hot state and does not enter the cool state.  
This behavior is similar to the first $\sim$60~d phenomenon 
of the 2021 anomalous event in SS Cyg (see also Figure 
\ref{lc-standstill}), though the amplitude of oscillations 
is a little smaller, and the duration of oscillations is 
a little shorter than observations.  

To reproduce the increment of the $V$-band 
magnitude in the first $\sim$60-d light curve of 
the 2021 anomalous event in SS Cyg against the normal quiescence, 
the inner disk should stay in the hot state during this event. 
It is unlikely that the WD or the secondary star or 
the bright spot becomes more than an order of magnitude brighter. 
The maximum brightness of the cool disk is estimated from 
the maximum effective temperature of the cool state to be 
less than 11.5 mag in the $V$ band, which is much fainter than 
the minimum $V$-band magnitude (= $\sim$10.5~mag) of 
the observed light curve during the first $\sim$60-d of 
the anomalous event (see Figure \ref{lc-standstill}). 
Our simulation displayed in Figure \ref{MTB-overflow} at least 
satisfies with the condition that the innermost part of 
the disk never returns to the cool state.

\ifnum0=1
\begin{figure}[htb]
\begin{center}
\FigureFile(80mm, 50mm){EVOFmodel-caseB-ah015-ac0015-to-012-f05-mdot169-rt0356-Nd002.ps}
\end{center}
\caption{
Results of our simulations in the case where $f_{\rm overflow}$ and $\alpha_{\rm cool} / \alpha_{\rm hot}$ simultaneously change on 240~d.  
Here, $f_{\rm overflow}$ is increased to 0.5 and $\alpha_{\rm cool} / \alpha_{\rm hot}$ is increased to 0.8 after 240~d.  
See the caption of Figure \ref{basic} as for the black and green lines in the third panel.  
}
\label{EV-overflow}
\end{figure}

Let us now return back to the case for the mass transfer 
rate of $\dot{M}_{\rm tr} = 10^{16.9}$~g~s$^{-1}$, i.e., 
that of our standard model. We find that in this case 
the inner disk drops to the cool state even if 
$f_{\rm overflow}$ is extremely high (see Figures 
\ref{overflow-increase} and \ref{overflow-model}).  
In order to reproduce the 2021 anomalous event of SS Cyg, 
we have this time attempted simulations by simultaneously 
changing $\alpha_{\rm cool}$ and $f_{\rm overflow}$ after 240~d.  
Figure \ref{EV-overflow} shows the simulation result where 
$\alpha_{\rm cool} / \alpha_{\rm hot}$ is switched from 
0.1 to 0.8 and $f_{\rm overflow}$ is switched from 0.0 to 0.5.  
By comparing this with the second and third panels of 
Figure \ref{overflow-vband} in which only $f_{\rm overflow}$ 
is amplified, we find that the quiescent state now disappears 
and the light curve shows small-amplitude fluctuations with 
a mid-brightness level.  
In the case that only $\alpha_{\rm cool}$ is amplified, 
the outburst amplitude is still $\sim$2 mag (see Figure 
\ref{EVmodel}).  However, the amplitude of fluctuations, 
in this case, is as low as 1 mag in Figure \ref{EV-overflow}.  
By incorporating both enhanced viscosity and gas-stream 
overflow, the inner part can easily keep the hot state.  
Moreover, it takes more time for the outer disk to enter 
an outburst, and the number of inside-out outbursts increases 
because the transferred mass is distributed to the disk 
outer rim and the region near $r_{\rm LS}$.  
As a result, the mid-brightness interval with low-amplitude 
oscillations lasts for a while.  
The light curve behavior after 240~d in Figure \ref{EV-overflow} 
is similar to the first $\sim$60~d behavior of the 2021 anomalous 
event in SS Cyg (see also Figure \ref{lc-standstill}). 
\fi

\section{Discussion}

\subsection{Does the 2021 anomalous event originate from the enhanced viscosity ?}

\citet{kim21sscyg} suggested that the enhancement of 
viscosity in the cool state would reproduce the 2021 
anomalous event and its forerunner in SS Cyg.  
We find that some of the light curves in our simulations of 
the enhanced viscosity model are similar to the observed 
light curve during the forerunner of the anomalous event.  
It is, however, difficult to reproduce the first $\sim$60~d 
behavior of the 2021 anomalous event because our simulated 
light curve still shows an outbursting behavior with 
amplitude as large as $\sim$2 mag even in the case of   
$\alpha_{\rm cool} / \alpha_{\rm hot}$ = 1.0, contrary to 
the observed light fluctuations with amplitude $\sim$1 mag.  
As discussed in section 3.2, the light curve behavior 
in the case of the same $\alpha$ for the hot and cool states 
may be sensitive to the S-shaped thermal-equilibrium curve 
used in the simulation. 
If one uses other S-shaped curves with a shorter intermediate 
branch, the outburst amplitude may become $\sim$1 mag 
but this is another problem.

The gas-stream overflow may play an important role 
to reproduce the 2021 anomalous event.  
When the stream overflow is taken into account in 
our simulations, small-amplitude inside-out outbursts 
begin to occur (see also Figures \ref{overflow-increase}, 
\ref{overflow-model}, and \ref{overflow-vband}).  
If the mass transfer rate is a little higher 
than that of our standard model in our simulations, 
the stream overflow generates a light curve similar to 
the anomalous event (see Figure \ref{MTB-overflow}).
The overflow is likely to have occurred at least during 
the first $\sim$60~d of the anomalous event.

\subsection{Can the gas-stream overflow explain the observed X-ray luminosity ?}

As pointed out in section 2, there is a large discrepancy 
between the observed X-ray luminosity in quiescence of 
SS Cyg and that predicted from the standard model 
in which the inner edge of the cold disk extends down 
near the WD surface.  
As discussed in section 2, we adopt the evaporation model, 
in which the geometrically thin viscous disk is truncated 
at the inner edge, $r_{\rm in}$, below which the disk 
expands to the hot spherical corona.  
In such cases, if $r_{\rm in}$ or 
$\alpha_{\rm cool}$ is increased, 
$\dot{M}_{\rm acc} (r_{\rm in})$ increases, which makes 
the predicted X-ray luminosity higher.  
However, there must be some limit to doing so.  
We have tried simulations where the inner edge of the disk is 
chosen to be $r_{\rm in} = 1.0 \times 
10^{10}$~cm by assuming that a large part of the inner disk is 
evaporated, and calculated the X-ray luminosity in 
quiescence to be $\sim$2$\times$10$^{32}$~erg~s$^{-1}$.  
However, this  X-ray luminosity is still lower than 
the observed one. 
It is noted here that the X-ray luminosity in quiescence 
estimated here depends on the thermal equilibrium curve used. 
The thermal equilibrium curve in this paper is based 
on the computations in \citet{min83DNDI}. 
If another thermal equilibrium curve such as that 
of \citet{ham98diskmodel} is used in which the convective 
energy transport is much more efficient in the vertical 
structure calculations, the X-ray luminosity of the cold disk 
could become larger, as large as a factor 3 of our estimate 
\citep{las08heliumdisk}.

The main difficulty in explaining fairly high X-ray flux 
observed in quiescence in SS Cyg lies in the difficulty 
in the first place to transfer a large amount of mass 
from the outer disk to the inner disk by viscous diffusion 
because of low viscosity in quiescence. 
To solve this difficulty, one possible solution is 
to consider a situation in that a part of the overflowing gas 
is directly transferred to the inner disk 
by the gas stream and to the inner edge of the disk from 
where the disk matter evaporates into the hot corona 
(i.e., a kind of shortcut in mass supply). However, 
we do not here discuss any specific mechanism for how 
the overflowing gas stream is converted there to 
the hot coronal matter.  We simply assume that the disk 
matter at the inner disk edge evaporates into the hot corona. 
For example, the observed X-ray luminosity during normal 
quiescence is reproduced if $\sim$5\% of the transferred 
mass is deposited directly to the inner edge and then to the hot corona via the gas-stream overflow.  
Even if $\sim$5\% of the transferred mass overflows 
the disk surface, the outburst behavior does not change 
so much in comparison with that of the standard model.  
The gas-stream overflow may alleviate the discrepancy 
between the observed and simulated X-ray luminosity.  

We next consider the X-ray luminosity during the 2021 
anomalous event.  
The light curve displayed in Figure \ref{MTB-overflow} 
is closest to the 2021 anomalous event in SS Cyg among 
our simulations.  
In this case, the inner disk always stays in the hot state, and 
the accretion rate at $r_{\rm in}$ during the mid-brightness 
interval after 240~d is as high as 
$\sim$4$\times$10$^{16}$~g~s$^{-1}$ on average.  
The X-ray luminosity is calculated to be  
$\sim$5$\times$10$^{33}$~erg~s$^{-1}$ by equation (\ref{xray-lumi}), 
which is a little higher but consistent with the observed X-ray luminosity. 
Although we have used the result in Figure \ref{MTB-overflow} 
in this calculation, the above discussion is not limited 
to this specific case in this simulation.  
The X-ray luminosity in the 2021 anomalous event would be 
reproduced by other simulations where the inner disk keeps 
the hot state and the averaged optical flux is as high as 
$\sim$10~mag because a large amount of mass accreted from 
the inner disk edge to the X-ray emitting hot corona.  
In this case, it is not required that the overflowing gas 
is directly injected into the hot corona to explain 
the observed X-ray luminosity.  
Nevertheless, the stream overflow is needed in our simulations 
to reproduce the light-curve behavior similar to 
the first $\sim$60~d light curve of the 2021 anomalous 
event in SS Cyg and to increase the mass accretion rate 
at the inner disk edge.

\section{Summary}

We have performed numerical simulations of 
the light curves for the 2021 anomalous event in SS Cyg 
by varying mass-transfer rates, varying viscosity 
in the cool state, and varying overflow rates of 
the gas stream.
Our findings are listed below.  

\begin{itemize}
\item If the mass-transfer rate increases, the cycle 
length of long outbursts becomes shorter initially.  
If the mass-transfer rate is enhanced further, 
the duration of long outbursts gradually lengthens, 
and the cycle length becomes longer. 
Finally, the long outburst persists all the way, 
and the system enters the hot steady state.
\item In the enhanced mass-transfer model, the outburst 
amplitude remains large even when the system approaches 
to the hot steady state, and we do not find any cases 
in which the outburst amplitude diminishes and 
finally vanishes with the increase in the mass-transfer rate.
\item It is found that when the mass transfer rate is gradually 
increased, the transition in light curves from the DN-type outburst 
state to the nova-like state occurs rather suddenly within 
a very narrow range in mass transfer rate; 
a condition favorable to explain the Z Cam phenomenon.
\item If the viscosity in the cool state is enhanced, 
inside-out outbursts frequently occur, and the quiescent 
level increases.
\item If $\alpha_{\rm cool}$ is chosen to be equal to 
$\alpha_{\rm hot}$, it is found that the inner disk never 
drops to the cool state for parameters corresponding to 
our standard model for SS Cyg. 
Even in this case, fairly large amplitude oscillatory 
light variations are found to occur with an amplitude of 2~mag, 
contrary to a naive expectation that small-amplitude 
fluctuations may result in such a case. 
\item We have studied the effect of gas-stream overflow 
beyond the outer disk rim in our light curve simulations where 
the mass by overflowing gas is assumed to be supplied 
to the inner disk around the circularization radius.  
If the overflow rate is increased in our simulations, 
a new type of inside-out type outbursts with an amplitude of 
1~mag occurs, and its number increases. 
It is found that any models with gas-stream overflow cannot 
reproduce the light curve of the 2021 anomalous event of SS Cyg
with parameters of our standard model.   
\item Within our simulations, 
only a model in which the gas-stream overflow is 
considered, and the mass transfer rate is a little higher 
than that of our standard model, may reproduce 
a light curve with fluctuations having amplitudes of 
less than 1 mag as observed in the 2021 anomalous event 
of SS Cyg.
\item The X-ray flux calculated from our simulations 
with the standard parameters is much lower than the observed 
X-ray flux during normal quiescence in SS Cyg. 
If a few percent of the overflowing gas is provided to 
the inner edge of the disk, the observed flux could be 
explained. 
On the other hand, the direct injection of the overflowing gas 
to the inner edge of the disk is not necessary to reproduce 
the observed X-ray flux in the case of the 2021 anomalous 
event since the accretion rate at the inner disk edge is 
high enough.
\end{itemize}

We conclude that the enhanced mass transfer cannot reproduce 
the 2021 anomalous event and its forerunner in SS Cyg.  
We have confirmed that the enhancement of viscosity 
in the cool state may reproduce the observational features 
of the forerunner of SS Cyg, as suggested in 
\citet{kim21sscyg}. However, it is not enough to 
reproduce the light-curve behavior of the first 60 days 
of the 2021 anomalous event.  
The gas-stream overflow may be necessary to explain this event.
Also, the stream overflow may play an important role 
in reproducing the observed X-ray flux during normal quiescence.

\section*{Acknowledgements}

M.~Kimura acknowledges support by the Special Postdoctoral 
ResearchersProgram at RIKEN.  
We are thankful to many amateur observers for providing 
a lot of data used in this research.  
This work was financially supported by Japan Society for 
the Promotion of Science Grants-in-Aid 
for Scientific Research (KAKENHI) Grant Numbers
JP20K22374 (MK) and JP21K13970 (MK).

\appendix

\section*{1.~~Details of the modifications of our numerical code}

In \citet{kim20tiltdiskmodel}, we treated the tilted disk 
and mixed logarithmic and linear meshes.  
In this study, we set $N_0$ concentric annuli from 
the innermost disk radius, $r_{\rm in}$, to the tidal 
truncation radius, $r_{\rm tidal}$, equally spaced 
in $\sqrt{r}$ as used in \citep{can93DI}.  
We first set $N_0$ to be 200 and the number of meshes ($N$) 
in the time-dependent calculation is variable.  
The center of each annulus is given by
\begin{eqnarray}
r_{\rm i - 1/2} &=& \frac{(\sqrt{r_{\rm in}} + \sqrt{dr} \times (i - 1))^2 + (\sqrt{r_{\rm in}} + \sqrt{dr} \times i)^2}{2}\\ 
{\rm for}~i &=& 1, 2, \dots, N, \\ \nonumber
\sqrt{dr} &=& \frac{\sqrt{r_{\rm tidal}} - \sqrt{r_{\rm in}}}{N_0}.
\label{mesh}
\end{eqnarray}

The mass supply pattern to the disk is also modified.  
We basically input the mass transferred from the secondary 
star to the outer disk edge with the width of $dr_{\rm s}$.  
The source term $s(r)$ is given by 
\begin{equation}
s(r) = \frac{\dot{M}_{\rm tr}}{dr_{\rm s}}, 
\label{source-term}
\end{equation}
where $\dot{M}_{\rm tr}$ represents the mass transfer rate.  
We input the heating by the energy dissipation of 
the gas stream from the secondary star to the outermost 
$N_{\rm S}$ meshes between $r_N - dr_{\rm s}$ and $r_N$.  
In this paper, $dr_{\rm s}$ is 0.02$a$ in the case of 
$N_0 = 200$, and a typical value of $N_{\rm S}$ 
is 7 in this paper.

\section*{2.~~Extra tidal heating due to the tidal truncation}

In our formulation, we assume that the tidal truncation 
radius works as a solid brick wall, that is, when 
the disk reaches the tidal truncation radius and tries to 
expand further, it is assumed that the expansion 
stops at the tidal truncation radius due to the strong 
tidal torques.  
As for the tidal torque, our formulation is basically 
the same as that in our previous paper \citep{kim20tiltdiskmodel}. 
The tidal torque is given by the following equation 
(see, equation (7) of \citet{kim20tiltdiskmodel}) 
if the disk radius is less than the tidal truncation radius: 
\begin{equation}
D = c\omega r \Sigma \nu \left(\frac{r}{a} \right)^5,  
\label{D-each}
\end{equation}
where $\nu$ is the kinematic viscosity (see, \citet{kim20tiltdiskmodel}).  
In this paper, we adopt the constant $c\omega$ to be 
0.4~s$^{-1}$, which is much smaller than that ($c\omega 
\simeq 9.5$~s$^{-1}$) expected in the case when the disk radius 
in the hot steady state just reaches the tidal truncation 
radius in this formulation. 
The tidal torque by equation (\ref{D-each}) is thus weak and 
we have to remove the extra angular momentum if the disk radius 
tries to exceed the tidal truncation radius.
In our previous paper \citep{kim20tiltdiskmodel}, 
the removal of the extra angular momentum at the tidal 
truncation radius was automatically taken into account, 
but the extra tidal heating due to this effect was 
not considered there.

Here we take into account this heating effect, and 
we implement in our code the extra tidal heating at 
the tidal truncation radius. To do so, we need to 
know the extra tidal torque by the tidal truncation 
(expressed here as $D_{\rm ext}$) in its explicit form. 
Once we know this extra tidal torque, we can easily 
calculate the extra tidal heating. 
Let us now consider a fictitious situation in which 
the tidal truncation were not exerted and so the disk were 
allowed to expand beyond $r_{\rm tidal}$ and to reach $r_N$, 
The total angular momentum of the disk is given by 
\begin{eqnarray}
J_{\rm disk}^{\rm new} - J_{\rm disk} &=& (h_{\rm LS} \dot{M}_{\rm tr} - h_0 \dot{M}_0 - D_{\rm total})dt \\ \nonumber
&=& \sum_{i-1}^{N-1} h_{i-1/2}^{\rm new} \Delta M_{i - 1/2}^{\rm new} + h_{N-1/2}^{\rm new} \Delta M_{N-1/2}^{\rm new}, 
\label{Jnew-disk-1}
\end{eqnarray}
where the quantities with and without the subscript `new' 
denote before and after a time step. 
Here, $h_{N-1/2}^{\rm new}$ is defined as 
$\sqrt{G M_1 (r_{N-1} + r_N)/2}$.
The other symbols appearing in equations (A4) 
were explained in Appendix 1 in \citet{kim20tiltdiskmodel}.

In our formulation, it is assumed that the disk radius 
is limited to the tidal truncation radius 
by the extra tidal torque when the disk 
tries to exceed the tidal truncation radius.  
In such a case, the conservation of the disk angular 
momentum given after a time step ($dt$) by equation (A4) 
is written as follows: 
\begin{eqnarray}
J_{\rm disk}^{\rm new'} - J_{\rm disk} &=& (h_{\rm LS} \dot{M}_{\rm tr} - h_0 \dot{M}_0 - D_{\rm total} - D_{\rm ext})dt \\ \nonumber
&=& \sum_{i-1}^{N-1} h_{i-1/2}^{\rm new} \Delta M_{i - 1/2}^{\rm new} + h_{N-1/2}^{\rm new'} \Delta M_{N-1/2}^{\rm new}, 
\label{Jnew-disk-2}
\end{eqnarray}
where $J_{\rm disk}^{\rm new'}$ and $h_{N-1/2}^{\rm new'}$ 
are the total angular momentum of the disk and the specific 
angular momentum of the largest annulus, respectively, 
in the case when the disk stops at the tidal truncation 
radius due to $D_{\rm ext}$. 
The quantity $h_{N-1/2}^{\rm new'}$ is given by 
$\sqrt{G M_1 (r_{N-1} + r_{\rm tidal})/2}$.

By comparing these two cases, we then obtain $D_{\rm ext}$ 
that is given by 
\begin{equation}
D_{\rm ext} = (J_{\rm disk}^{\rm new} - J_{\rm disk}^{\rm new'})/dt = ((h_{N-1/2}^{\rm new} - h_{N-1/2}^{\rm new'}) \Delta M_{N-1/2}^{\rm new})/dt.  
\label{Dext}
\end{equation}
That is, the extra tidal torque, $D_{\rm ext}$, is 
obtained from a fictitious calculation in which 
the disk is allowed to expand beyond the tidal 
truncation radius.

\begin{figure}[htb]
\begin{center}
\FigureFile(80mm, 50mm){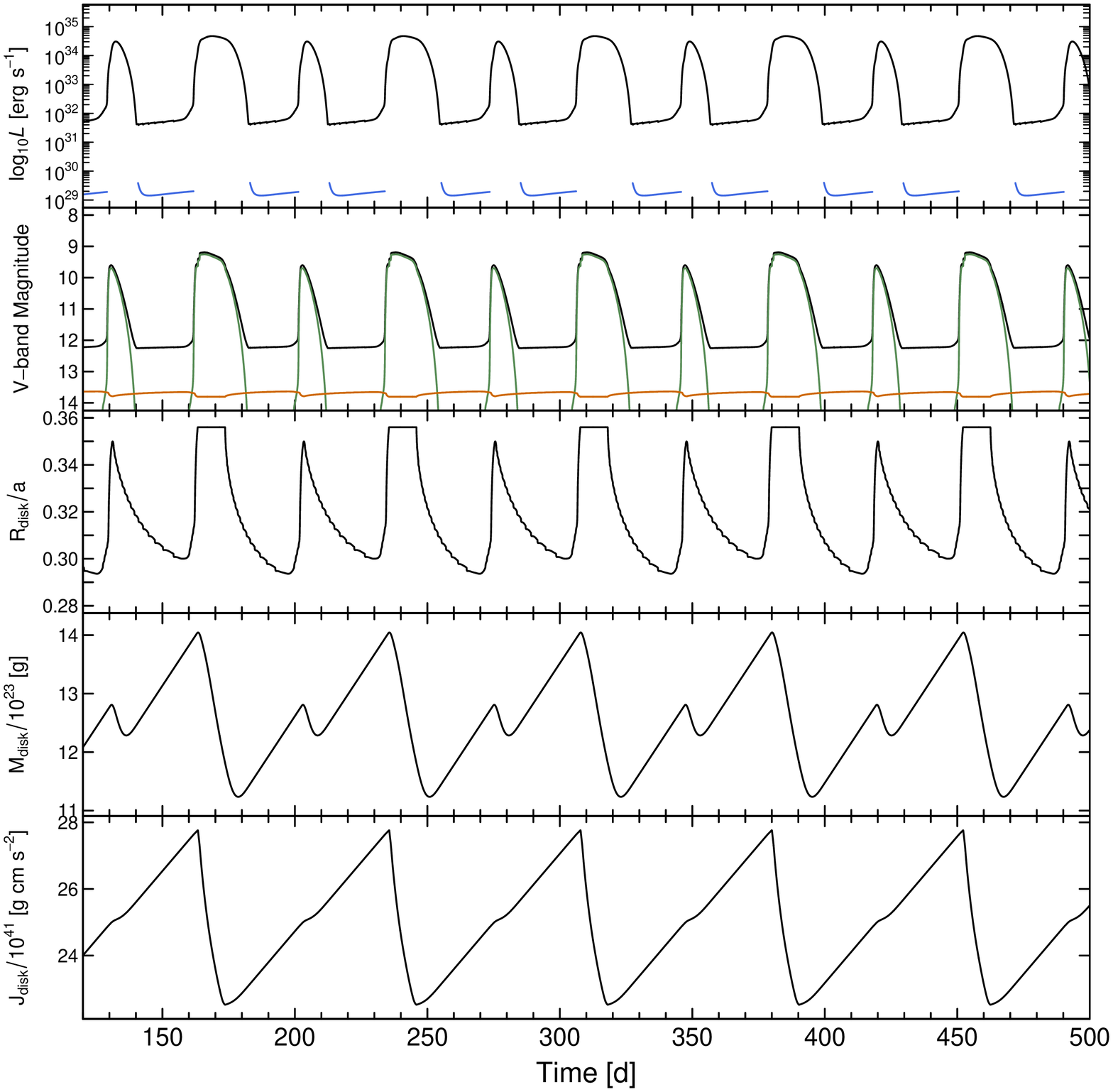}
\end{center}
\caption{
Same as Figure \ref{basic} but in the case of $dr_{\rm d} = 0.005a$.  
}
\label{basic-Wd-small}
\end{figure}

We now calculate the extra tidal heating by assuming 
that it is exerted in the outermost region of the disk 
with the width of $dr_{\rm d}$. 
We set $dr_{\rm d}$ to be 0.02$a$ in this study.  
If we introduce a quantity, $d_{\rm ext}$, which is given by  
\begin{equation}
d_{\rm ext} = \frac{D_{\rm ext}}{dr_{\rm d}},
\label{dext}
\end{equation}
We then add the dissipation rate by the extra 
tidal torque to the outermost 
$N_{\rm d}$ meshes between $r_{N} - dr_{\rm d}$ and $r_{N}$ by 
\begin{equation}
Q_{\rm 2ext, i - 1/2} = d_{\rm ext} \frac{\Omega_{i - 1/2} - \omega_{\rm orb}}{2 \pi r_{i - 1/2}}, 
\label{dext}
\end{equation}
where $\Omega_{i - 1/2}$ is the Keplerian angular velocity 
at $r_{i-1/2}$ written by $\sqrt{G M_1 {r_{i-1/2}^3}}$ and 
$\omega_{\rm orb}$ is the orbital angular velocity, 
respectively.

We may note here that $dr_{\rm d}$ affects the number of 
short outbursts sandwiched by two long outbursts.  
As $dr_{\rm d}$ becomes smaller, the number of short outbursts 
between two long outbursts becomes smaller.  
This is because an outburst easily enters a long plateau 
stage with smaller $dr_{\rm d}$ because of strong tidal torques 
at the outer disk edge when the disk radius reaches the tidal 
truncation radius.  
We exhibits an example of light curves in the case of 
$dr_{\rm d} = 0.005a$ in Figure \ref{basic-Wd-small}.


\newcommand{\noop}[1]{}

\end{document}